\documentclass[prd,nobibnotes,onecolumn, nofootinbib]{revtex4}
\usepackage{graphics,color,array,dcolumn}
\usepackage{calc}

\usepackage{amsmath}
\usepackage{amssymb}
\usepackage{xspace}

\newlength{\figurewidth}
\allowdisplaybreaks[1]

\begin{document}
\setlength{\figurewidth}{\columnwidth}

\title{Renormalization Group Flow in Scalar-Tensor Theories. I}
\author{Gaurav Narain}
\affiliation{SISSA, Via Beirut 4, 34151 Trieste, Italy}
\email{narain@sissa.it}
\author{Roberto Percacci\footnote{\it on leave from SISSA, via Beirut 4, I-34151 Trieste, Italy. 
Supported in part by INFN, Sezione di Trieste, Italy}}
\email{rpercacci@perimeterinstitute.ca}
\affiliation{Perimeter Institute for Theoretical Physics, 31 Caroline St. North, Waterloo, Ontario N2J 2Y5, 
Canada}

\begin{abstract}
We study the renormalization group flow in a class of scalar-tensor theories involving 
at most two derivatives of the fields.
We show in general that minimal coupling is self consistent, in the sense that when the 
scalar self couplings are switched off, their beta functions also vanish.
Complete, explicit beta functions that could be applied to a variety of cosmological models
are given in a five parameter truncation of the theory in $d=4$.
In any dimension $d>2$ we find that the flow has only a ``Gaussian Matter'' fixed point,
where all scalar self interactions vanish but Newton's constant and the cosmological constant are nontrivial. 
The properties of these fixed points can be studied algebraically to some extent.
In $d=3$ we also find a gravitationally dressed version of the Wilson-Fisher fixed point, 
but it seems to have unphysical properties.
These findings are in accordance with the hypothesis that these theories are asymptotically safe.
\end{abstract} 

\maketitle

\section{Introduction}

Fundamental scalar fields have not yet been observed,
but they play a crucial role in the standard model and in grand unified theories,
as the order parameters whose VEV is used to distinguish between otherwise
undifferentiated gauge interactions.
Whether such scalar order parameters are elementary fields, as in the standard model,
or composites, as in technicolor theories, is still an open question.
Known examples of the Higgs phenomenon (superconductivity, 
the chiral condensate in QCD) point to the latter possibility, 
but even if this was the case it might still be possible to
use scalar theory as an effective description (\'a la Landau-Ginzburg)
at sufficiently low energy.

Scalar fields also play an important role in theories of gravity.
Due to their simplicity they are very often used as models for matter.
Also, because of the ease by which one can generate a nontrivial VEV,
with an energy momentum tensor that resembles a cosmological constant,
a scalar field is the most popular option as a driver of inflation.
Furthermore, scalar fields easily mingle with the metric: 
by means of Weyl transformations
it is possible to rewrite the dynamics in different ways \cite{Scalartensor},
sometimes leading to new insight or to simplifications.
Theories of gravity in which a scalar is present
are often called scalar-tensor theories.
In this paper we will discuss the quantum properties of a class of theories
of this type.

The original motivation for this work comes from the progress that
has been made in recent years towards understanding the UV behaviour of gravity. 
It seems that pure gravity possesses a
Fixed Point (FP) with the right properties to make it asymptotically safe,
or in other words nonperturbatively renormalizable
\cite{Weinberg, Dou, Reuter, Souma, Lauscher, Litim:ed, 
Weyer, Codello, CPR, PercacciN, Lauscher2, MachSau, frank2, BMS, 
crehroberto, creh1, elisa, AGS,Niedermaier:2009zz} (see also \cite{AS_rev} for reviews).
Let us assume for a moment that this ambitious goal can be achieved,
and that pure gravity can be shown to be asymptotically safe.
Still, from the point of view of phenomenology, we could not be satisfied
because the real world contains also dozens of matter fields that interact
in other ways than gravitationally, and a their presence affects also
the quantum properties of the gravitational field, as is known since long \cite{veltman}.
Indeed, in a first investigation along these lines, it was shown in \cite{Perini1}
that the presence of minimally coupled ({\it i.e.} non self interacting) matter fields
shifts the position of the gravitational FP and the corresponding critical exponents.
In some cases the FP ceases to exist, so it was suggested that this
could be used to put bounds on the number of matter fields of each spin.
More generally the asymptotic safety program requires that the fully interacting theory
of gravity and matter has a FP with the right properties.
Given the bewildering number of possibilities, in the search for such a theory
one needs some guiding principle.
One possibility that naturally suggests itself is that all matter self-interactions
are asymptotically free \cite{fradkintseytlin}.
Then, asymptotic safety requires the existence of a FP where the matter couplings
approach zero in the UV, while the gravitational sector remains interacting.
We will call such a FP a ``Gaussian Matter FP'' or GMFP.
Following a time honored tradition, as a first step in this direction, 
scalar self interactions have been studied in \cite{GrigPerc,Perini2}.
Here we pursue that study further.

The tool that we use is the Wetterich equation, an exact renormalization group (RG) flow equation 
for a type of Wilsonian effective action $\Gamma_k$, called the ``effective average action''.
This functional, depending on an external energy scale $k$, 
can be formally defined by introducing an IR suppression in the functional integral
for the modes with momenta lower than $k$.
This amounts to modifying the propagator of all fields, leaving the interactions untouched.
Then one can obtain a simple functional RG equation (FRGE) for the dependence of $\Gamma_k$
on $k$ \cite{ Wetterich, Morris, Bagnuls, Berges:2000ew, Gies,  Pawlowski}. Insofar as 
the effective average action contains information about all the couplings in the theory, 
this functional RG equation contains all the beta functions of the theory.
In certain approximations one can use this equation to reproduce the
one loop beta functions, but in principle the information one can
extract from it is nonperturbative, in the sense that is does not
depend on the couplings being small.

The most common way of approximating the FRGE is to do derivative expansion of 
effective average action and truncate it at some order.
In the case of a scalar theory the lowest order of this expansion is the
local potential approximation (LPA), where one retains a standard kinetic term
plus a generic potential \cite{Morris, Wetterich, Morris:1994ki,LitimBer}.
In the case of pure gravity, the derivative expansion
involves operators that are powers of curvatures and derivatives thereof.
This has been studied systematically up to terms with four derivatives
\cite{Codello, BMS,Niedermaier:2009zz}
and for a limited class of operators (namely powers of the scalar curvature)
up to sixteen derivatives of the metric \cite{CPR, MachSau}.
In the case of scalar tensor theories of gravity, one will have to
expand both in derivatives of the metric and of the scalar field.

In this paper we study (Euclidean) effective average actions of the form
\begin{equation}
\label{eq:action}
\Gamma_k [g,\phi ] = \int\,{d}^{d} x\,\sqrt{g}\left(V(\phi^2)-F(\phi^2)R
+\frac{1}{2}  g^{\mu \nu} \partial_{\mu} \phi \partial_{\nu} \phi \right)
+S_{GF}+S_{gh}.
\end{equation}
This can be seen as a generalization of the LPA, where one also
includes terms with two derivatives of the metric.

In \cite{Perini2} it was shown that in $d=4$, and assuming that
$V$ and $F$ are polynomials in $\phi^2$, this theory admits a
GMFP where only the lowest
($\phi^2$-independent) coefficient in $V$ and $F$ are nonzero.
In this paper we extend and generalize this result in various ways.
First of all, using so called ``optimized'' cutoff types \cite{Litim:cf}
it is possible to write the beta functions in closed form, 
whereas in \cite{Perini2} they could only be studied numerically.
This makes the subsequent analyses much more transparent.
Unlike in \cite{Perini2}, we will not assume from the outset that $V$
and $F$ are polynomials. Then, using the optimized cutoff it is possible
to write explicit beta functionals for $V$ and $F$.
Exploiting general properties of these functionals
we will be able to prove certain properties
of the linearized flow in the neighborhood of the FP
which had only been numerically observed previously.
Namely we show that the matrix describing the linearized flow only has
nonzero entries on the diagonal and on three lines next to it,
and furthermore it has a block structure such that knowledge of the
first two $2\times2$ blocks determines the whole matrix.

The discussion in this paper is also more general than that of \cite{Perini2}
in two ways: we keep the dimension of spacetime arbitrary
and we allow for a more general gauge fixing, depending on
two arbitrary parameters.
Keeping the dimension general is useful in view of possible 
applications to popular ``large extra dimensions'' theories \cite{Litim:ed},
and also to higher dimensional dilatation-symmetric models which lead
to vanishing cosmological constant in four dimensions \cite{wetterichcosmology}.
Furthermore, with the closed form beta functions we can also perform a better search for
other FP's where the scalar interactions are not all turned off.
In \cite{Perini2} a numerical search was conducted on a grid of points
in the neighborhood of the Gaussian FP, and no nontrivial FPs were found.
Here, having the closed form of the beta functions, we can look for FPs
by different methods. In $d=4$, $5$ and $6$, some such points are found, but they appear to
be spurious.
On the other hand in $d=3$ there is a FP which seems to be a genuine generalization
of the Wilson-Fischer FP \cite{WF}, but it has unphysical properties.

In a companion paper \cite{gau_chris} we will extend the discussion to a more general
class of effective actions, 
\begin{equation}
\label{eq:action2}
\Gamma [g,\phi ] = \int\,{d}^{d} x \, \sqrt{g} \, \left(L(\phi^2,R)
+\frac{1}{2} \,  g^{\mu \nu}  \, \partial_{\mu} \phi \partial_{\nu} \phi \right)
+S_{GF}  + S_{gh}.
\end{equation}
where $L$ is in general a function of the curvature scalar and of the scalar field.

Even though the original motivation of our work was to study the UV
properties of the theory, it is important to stress that the beta functions
that we obtain are completely general: they hold for any energy range.
Depending on the ratio between the parameters of the theory (the cosmological constant,
Newton's constant, the scalar mass and all the dimensionful higher couplings)
different terms in the beta functions will come to dominate.
However, this is something that has not been put in a priori.
Thus the beta functions can be used also to study IR or mesoscopic problems,
provided the system can be accurately modelled by a scalar tensor theory.

There is clearly much scope for applications to cosmology.
Early work in this direction has been done in \cite{Reutercosmology},
using the beta functions of pure gravity.
Along a different line, given the role played by scalar fields in inflation
it seems likely that the RG running of couplings could have significant effects.
This seems particularly true of recent attempts to use the standard model Higgs
field as an inflaton, which use a special case of the action \eqref{eq:action}
$$
\sqrt{g}\left(
\frac{1}{2} g^{\mu \nu} \partial_{\mu} \phi \partial_{\nu} \phi 
+\frac{1}{2}m^2\phi^2+\lambda\phi^4+\frac{1}{16\pi^2}(2\Lambda-R)+\frac{1}{2}\xi\phi^2
\right)
$$
with a large value for the nonminimal coupling $\xi$ \cite{higgsflaton}.
The beta functions given in Appendix A, contain the full dependence
on $\Lambda$, $G$, $m^2$, $\lambda$ and $\xi$, including threshold effects
and a resummation of infinitely many perturbative contributions.

One can imagine also applications in the IR,
for example along the lines of \cite{wetterichcosmology}.
We mention that the appearance of a scalar field in the low energy description of gravity
has been also stressed in \cite{mottola}.
For a FRGE-based approach to that issue see also \cite{crehroberto}.

This paper is organized as follows. In section 2 we will derive the ``beta
functionals'' for $V$ and $F$.
In section 3 we discuss the general properties of the GMFP, in any gauge and dimension.
In section 4 we discuss numerical solutions for the GMFP.
In section 5 we will discuss other FP's with nontrivial potentials
and we conclude in section VI with some additional remarks.
Apendix A contains some lengthy formulae for the beta functions
of five couplings in four dimensions.

\section{The beta functions}

In this paper we will obtain beta functionals for the functions $V$ and $F$ defined in (\ref{eq:action}).
To achieve this, we use Wetterich's functional renormalization group equation (FRGE) \cite{Wetterich}
\begin{equation}
\label{eq:FRGE}
\partial _t \Gamma _k = \frac{1}{2}\mathrm{STr}
\Biggl[
\Bigg(\frac{\delta^2 \Gamma_k}{\delta \Phi \, \delta \Phi} + \mathcal{R}_k \Biggr)^{-1} \partial_t \mathcal{R}_k
 \Biggr] ,
\end{equation}
where $\Phi$ are all the fields present in the theory
and $\mathrm{STr}$ is the generalized functional trace 
including a minus sign for fermionic variables and a factor 2 for complex variables,
and $\mathcal{R}_k$ is a suitable tensorial cutoff.

\subsection{Second variations}

In order to evaluate the r.h.s. of (\ref{eq:FRGE}) we start from the second functional
derivatives of the functional (\ref{eq:action2}).
These can be obtained by expanding the action to second order in the quantum fields
around classical backgrounds: $g_{\mu\nu}=\bar g_{\mu\nu}+h_{\mu\nu}$ and
$\phi=\bar\phi+\delta\phi$, where $\bar\phi$ is constant. The gauge fixing action
is given by
\begin{gather}
\label{eq:GFaction}
S_{GF} =\frac{1}{2\alpha} \int \,{d}^{d} x \, \sqrt{\bar{g}} \, 
F(\phi^2)\bar{g}^{\mu \nu} \, \chi_{\mu} \, \chi_{\nu} \, 
\ , \\
\chi^{\mu}=\left(\bar{\nabla}_{\nu}h^{\nu \mu}-\frac{\beta+1}{d}\bar{\nabla}^{\mu}h\right) \, \ . \notag
\end{gather}
and $S_{gh}$ is the corresponding ghost action given by
\begin{equation}
\label{eq:GHaction}
S_{GH} = -  \int \, {d}^{d} x \, \sqrt{\bar{g}} \, \bar{C}^{\mu} \, \left[{\delta}^{\rho}_{\mu} \, \bar{\Box}  
+\left(1-\frac{2(1+\beta)}{d}\right) \bar{\nabla}_{\mu} \bar{\nabla}^{\rho}
+ \bar{R}^{\rho}_{\mu} \right]C_{\rho} \, \ . 
\end{equation}
These terms are already quadratic in the quantum fields.
The second variation of eq. (\ref{eq:action}) is,
\begin{eqnarray}
\label{eq:actexp}
\Gamma^{(2)}_k &=&
\frac{1}{2} \int \, {d}^{d} x  \sqrt{g} 
\Biggl[ 
\left( \frac{1}{4} h^2 - \frac{1}{2} h_{\mu \nu} h^{\mu \nu} \right ) \, \left( V(\phi^2) - F(\phi^2) \, R \right ) 
+ F(\phi^2)  \, h \, h^{\mu \nu} \, R_{\mu \nu} + \frac{1}{2} F(\phi^2) \, h \, \Box \, h 
\notag \\
&& 
+ F(\phi^2)  \, h^{\mu \nu } \nabla_{\mu} \nabla_{\rho} h^{\rho}{}_{\nu} 
- F(\phi^2)  \, h_{\alpha}{}^{\nu} \, h^{\mu \alpha} \, R_{\mu \nu } 
- F(\phi^2)  \, h^{\mu \nu} \, R_{\rho \mu \sigma  \nu } \, h^{\rho \sigma }
- \frac{1}{2} F(\phi^2)  h^{\mu \nu} \, \Box \, h_{\mu \nu} 
- F(\phi^2) \, h \nabla_{\mu} \nabla_{\nu} h^{\mu \nu} 
\Biggr] \notag \\
&& + \int \, {d}^{d} x  \sqrt{g} 
\Biggl[
h \cdot \phi \left( V^{\prime} (\phi^2) - F^{\prime} (\phi^2) \, R \right ) \, \delta \phi 
+ 2 \, \phi  F^{\prime} (\phi^2) \, h^{\mu \nu} \, R_{\mu \nu} \, \delta \phi 
- 2 \, \phi \,  F^{\prime} (\phi^2) \, \delta \phi \left( \nabla_{\mu} \nabla_{\nu} h^{\mu \nu} - \Box h \right) 
\Biggr] \notag \\
&& +  \frac{1}{2} \int \, {d}^{d} x  \sqrt{g} \, \delta\phi
\Biggl[ - \Box + 2\, V^{\prime} (\phi^2) + 4 \, \phi^2 \, V^{\prime \prime} (\phi^2)  
 - R \, \left(  2\, F^{\prime} (\phi^2) + 4 \, \phi^2 \, F^{\prime \prime} (\phi^2)  \right)
\Biggr] 
+ S_{GF} + S_{gh}
\, \ .
\end{eqnarray}
Since we will never have to deal with the original metric $g_{\mu\nu}$ and scalar field $\phi$,
in order to simplify the notation, in the preceding
formula and everywhere else from now on we will remove the bars from the backgrounds.
As explained in detail in \cite{Reuter}, the functional that obeys the FRGE (\ref{eq:FRGE})
depends separately on the background field $\bar g_{\mu\nu}$ and on a ``classical field'' 
$\left( g_{\mathrm{cl}} \right) _{\mu\nu}=\bar g_{\mu\nu}+ \left( h_{\mathrm{cl}} \right)_{\mu\nu}$,
where $ \left( h_{\mathrm{cl}} \right)_{\mu\nu}$ is Legendre conjugate to the sources that couple 
linearly to $h_{\mu\nu}$. The same applies to the scalar field.
In this paper, like in most of the literature on the subject, we will restrict ourselves
to studying the effective average action in
the case when $\left( g_{\mathrm{cl}} \right) _{\mu\nu}=\bar g_{\mu\nu}$ and $\phi_{\mathrm{cl}}=\bar\phi$.
From now on the notation $g_{\mu\nu}$ and $\phi$ will be used to denote equivalently 
the ``classical fields'' or the background fields.
For a discussion of the effective average action of pure gravity in the more general case
when $\left( g_{\mathrm{cl}} \right) _{\mu\nu}\not=\bar g_{\mu\nu}$
we refer to \cite{reutermanrique}.

\subsection{Decomposition}

In order to partially diagonalize the kinetic operator, 
we use the decomposition of $h_{\mu\nu}$ into irreducible components 
\begin{equation}
\label{eq:TTdecomp}
h_{\mu \nu} = h_{\mu \nu}^T + \nabla _{\mu} \xi _{\nu} + \nabla _{\nu} \xi _{\mu} + \nabla _{\mu} \nabla _{\nu} \sigma  - \frac{1}{d} g_{\mu \nu} \Box \sigma + \frac{1}{d} g_{\mu \nu} h \, \ ,
\end{equation}
where $h_{\mu \nu}^{T}$ is the (spin 2) transverse and traceless tensor, 
$\xi_{\mu}$ is the (spin 1) transverse vector component, 
$\sigma$ and $h$ are (spin 0) scalars. 
In some cases this decomposition allows an exact inversion of the propagator. 
This happens for example in the case of maximally symmetric background metric. 
Thus with that in mind we will work on a $d$-dimensional sphere. 
This change of variables in the functional integral gives rise to 
Jacobian determinants, which however can be absorbed by further field re-definitions 
$\hat\xi_\mu=\sqrt{-\Box-\frac{R}{d}} \, \xi_\mu$ and  
$\hat\sigma=\sqrt{-\Box}\sqrt{-\Box-\frac{R}{d-1}} \, \sigma$
\cite{Dou, Lauscher, CPR}. 
Then the inverse propagators 
for various components of the field are easily read from the second variation 
of the effective action. 
Thus for the spin-2 component $h_{\mu \nu}^T$ we get the following inverse propagator:
\begin{equation}
\label{eq:sp2propgrav}
\frac{1}{2} F(\phi^2)\left(-\Box+\frac{d^2-3d+4}{d(d-1)} \, R\right)
-\frac{1}{2}V(\phi^2) \, . 
\end{equation}
For the spin-1 component $\hat\xi$ we have the following inverse propagator:
\begin{equation}
\label{eq:sp1propgrav}
\frac{1}{\alpha}F(\phi^2)\left(-\Box-\frac{R}{d}\right)-V(\phi^2)
+\frac{d-2}{d} \, F(\phi^2) \, R \, .
\end{equation}
The two spin-0 components of the metric, $\hat\sigma$ and $h$, mix with the fluctuation of $\phi$ 
resulting in an inverse propagator 
given by a symmetric $3\times 3$ matrix $S$, with the following entries:
\begin{gather}
\label{eq:sp0propgrav}
S_{\sigma\sigma} 
= \left(1-\frac{1}{d} \right)   
\left[
\left\{ \frac{1}{2}-\left(1-\frac{1}{\alpha}\right)\left(1-\frac{1}{d}\right) \right\} \,
F(\phi^2) \, (-\Box) 
-\frac{1}{2} V(\phi^2 )+\frac{d-4}{2d} \, F(\phi^2) \, R
+\left(1-\frac{1}{\alpha}\right)F(\phi^2) \, \frac{R}{d} \right] \, \ , \notag \\
S_{\sigma h}=S_{h\sigma}
= \frac{1}{2}\left(1-\frac{1}{d} \right) \left[ \frac{2}{d} \left( \frac{\beta}{\alpha} + 1 \right)  - 1 \right] F(\phi^2) \, \sqrt{-\Box\left(-\Box-\frac{R}{d-1}\right)}\, \ , \notag \\
S_{\sigma\phi}=S_{\phi\sigma}=
-2 \, \phi \, \left(1-\frac{1}{d}\right) \, F'(\phi^2) \, \sqrt{-\Box\left(-\Box-\frac{R}{d-1}\right) }\, , \notag \\
S_{hh} = \frac{d-2}{4d}\left[
\left\{ -\left(1-\frac{1}{d}\right)+\frac{2\beta^2}{\alpha d(d-2)} \right \} \, 2 \, F(\phi^2) \, (-\Box) 
+V(\phi^2)-\frac{d-4}{d}F(\phi^2) \, R\right] \, \ ,  \notag \\
S_{h\phi}=S_{\phi h}= \left[ -2\left(1-\frac{1}{d}\right) \, \phi \,  F'(\phi^2) \, (-\Box)
+\phi \, V'(\phi^2)
-\left(1-\frac{2}{d}\right) \, \phi \, F'(\phi^2) \, R  \right]
\, \ , \notag \\
S_{\phi\phi} = -\Box+2V'({\phi}^2)+4\phi^2V''(\phi^2)-(2F'(\phi^2)
+4\phi^2F''(\phi^2))R \, .
\end{gather}

In order to diagonalize the kinetic operator occuring in the ghost action eq. (\ref{eq:GHaction}), 
we perform a similar decomposition of the ghost field into transverse and longitudinal parts in the following manner:
\begin{equation}
\label{eq:ghbreak}
\bar{C}^{\mu} = \bar{C}^{\mu T} + \nabla ^{\mu} \bar{C} \, \ , \qquad
C_{\mu} = C^{T}_{\mu} + \nabla_{\mu} C \, . 
\end{equation}
where $\bar{C}^{\mu T}$ and $C^{T}_{\mu}$ satisfy the following constraints,
\begin{gather}
\label{eq:ghconst}
\nabla_{\mu} \bar{C}^{\mu T} = 0 \, \ , \qquad  \nabla^{\mu} C^{T}_{\mu} =0.
\end{gather}
Again this decomposition would give rise to a non trivial Jacobian in the path-integral, which is cancelled by the further redefinition $\hat C=\sqrt{-\Box}C$.
For spin-1 component of the ghost, the inverse propagator is 
\begin{equation}
\label{eq:sp1propgh}
-\Box-\frac{R}{d} \, \ ,
\end{equation}
while for spin-0 component we have the following inverse propagator
\begin{equation}
\label{eq:sp0propgh}
\left(2-\frac{2(1+\beta)}{d}\right)(-\Box)-\frac{2R}{d} \, \ ,
\end{equation}

Now we have to specify the cutoff $\mathcal{R}_k$ occuring in FRGE eq. (\ref{eq:FRGE}).
We define $\mathcal{R}_k$ by the rule that $\Gamma^{(2)}_k+\mathcal{R}_k$ has the
same form as $\Gamma^{(2)}_k$ except for the replacement of $-\Box$ by $P_k(-\Box)$,
where $P_k(z) = z + R_k(z)$. $R_k(z)$ is a profile function which  tends to $k^2$ for $z\to 0$ and it approaches 
zero rapidly for $z>k^2$. The quantity $\Gamma^{(2)}_k+\mathcal{R}_k$ is the 
``modified inverse propagator''.
This procedure applies both to the bosonic degrees of freedom and to the ghosts.
The cutoff $\mathcal{R}_k$ occuring in the FRGE depends on $k$ not only 
through the profile function $R_k(z)$, but also through $k$ dependent couplings 
present in the function $F(\phi^2)$ and $F^{\prime} (\phi^2)$. 
Thus the derivative $k\frac{d}{dk}=\frac{d}{dt}$ acts not only on the profile function  
$R_k(z)$, but also on the $k$-dependent couplings present in $F(\phi ^2)$ and $F^{\prime} (\phi^2)$. 
When this is neglected one recovers the one loop results.
The presence of the beta functions on the RHS of the FRGE, 
produces a coupled system of linear equations,
which has to be solved algebraically to yield the beta functions.

\subsection{The $\beta$-functionals in $d=4$}

To read off the beta functions we have to compare the r.h.s. of the FRGE
with the $t$-derivative of eq. (\ref{eq:action}), namely
\begin{equation}
\partial_t\Gamma [g,\phi ] = \int\,{d}^{d} x \, \sqrt{g}
\left( \partial_t V(\phi^2,R)-\partial_t F \, R \right) \, .
\end{equation}
(the gauge fixing and the kinetic term are not allowed to run in our approximations).
Since the background $R$ and $\phi$ are constant, the space-time integral produces 
just a volume factor, which eventually cancels with the same factor appearing on the RHS 
of the FRGE. Thus the running of $V$ and $F$ can be calculated using,
\begin{gather}
\label{eq:betaVF_def}
\partial_{t}V(\phi^2) = \left. \frac{1}{Vol} \, \partial_{t}\Gamma_k\right |_{R=0}\ ,\hspace{10mm} 
\partial_{t}F(\phi^2) = \left. - \frac{1}{Vol} \, \frac{\partial(\partial_{t} \Gamma_k )}{\partial R}\right |_{R=0} \, \ ,
\end{gather}
where $Vol$ is the space-time volume. In order to exhibit the explicit form of these beta functionals 
we go to $d=4$, where $Vol=\frac{384\pi^2}{R^2}$, and set $\alpha=0$ and $\beta=1$ (De-Donder gauge). 
Furthermore, we choose the optimized cutoff
$R_k(z) = (k^{2} - z) \theta (k^{2} -z )$ \cite{Litim:cf},
which allows to perform the integrations 
occuring in FRGE trace in closed form (see appendix A in \cite{CPR}).
From the FRGE we then get
\begin{eqnarray}
\label{eq:betaVF1}
\partial_t V &=&
\frac{k^4}{192\pi^2}
\Biggl\{
6 + \frac{30 \, V}{\Psi} + \frac{6 (  k^2 \, \Psi + 24 \, \phi^2 \, k^2 \, F^{\prime} \, \Psi^{\prime} + k^2 \, F \Sigma_1) }{\Delta} 
+ \left( \frac{4}{F} + \frac{5 \, k^2}{\Psi} + \frac{ k^2 \, \Sigma_1}{\Delta} \right)  \partial_t F  
+ \frac{ 24 \, \phi^2 \, k^2  \, \Psi^{\prime}}{ \Delta} \, \partial_t F^{\prime} 
\Biggr\} \, \ , \\
\label{eq:betaVF2}
\partial_t F &=&
\frac{k^2}{2304\pi^2}
\Biggl\{ 
150 + \frac{120 \, k^2 \, F \, ( 3 \, k^2 F - V )}{\Psi ^2} 
- \frac{24}{\Delta} \left( 24 \, \phi ^2 \, k^2 \, F^{\prime} \, \Psi^{\prime} + k^2 \, \Psi +  k^2 \, F \Sigma_1 \right) 
 - \frac{36}{\Delta ^2} 
\Biggl[
 - 4 \, \phi^2 \, (6 \, k^4 \, F^{\prime 2} + \Psi^{\prime 2} ) \, \Delta  \notag \\
&&\!\!\!\!\!\!\!\!\!\!\!\!
+ 4 \, \phi^2 \, \Psi \, \Psi^{\prime} \, ( 7 \, k^2 \, F^{\prime} - V^{\prime} ) \, ( \Sigma _1 - k^2) 
+ 4 \, \phi^2 \Sigma _1 \, ( 7 \, k^2 \, F^{\prime} - V^{\prime} ) \, ( 2 \, \Psi \, V^{\prime} - V \, \Psi^{\prime} ) 
+ 2 \, k^4 \, \Psi^2 \, \Sigma_2  
+ 48 \, k^4 \, F^{\prime} \, \phi^2 \, \Psi \, \Psi^{\prime} \, \Sigma_2 \notag \\
&&\!\!\!\!\!\!\!\!\!\!\!\!
- 24 \, k^4 F \, \phi^2 \, \Psi^{\prime 2} \, \Sigma_2
\Biggr] 
- \frac{\partial_t F}{F}
\Biggl[
30  -  \frac{10 \, k^2F \, (7 \, \Psi + 4 \, V ) }{\Psi^2} 
+ \frac{6}{\Delta ^2}
\Biggl(
 k^2 \, F \, \Sigma_1 \, \Delta 
+ 4 \, \phi^2 \, V^{\prime} \, \Psi^{\prime} \, \Delta   -24 \, k^4 \, F \, \phi^2 \, \Psi^{\prime 2} \, \Sigma_2 \notag \\
&&\!\!\!\!\!\!\!\!\!\!\!\!
-4 \, \phi^2 \, k^2 \, F \, \Psi^{\prime} \, \Sigma_1 (7 \, k^2 \, F^{\prime}-V^{\prime})
\Biggr)
\Biggr]
+ \partial_t F^{\prime} \, \frac{24 \, k^2 \, \phi^2 }{\Delta ^2} \Biggl[
(  k^2 \, F^{\prime} + 5 \, V^{\prime} ) \Delta  - 12 \, k^2 \, \Psi \, \Psi^{\prime} \, \Sigma_2 
- 2 \, ( 7 \, k^2 F^{\prime} - V^{\prime} ) \, \Psi \, \Sigma_1 
\Biggr]
\Biggr\}
\end{eqnarray}
where we have defined the shorthands:
$$
\Psi = k^2 \, F - V \,  ;\quad
\Sigma_1 = k^2+ 2 \, V^{\prime} + 4 \, \phi^2 \, V^{\prime \prime} \, \ ;\quad
\Sigma_2 = 2 \, F^{\prime} + 4 \, \phi^2 \, F^{\prime \prime} \, \ ;\quad
\Delta = \left( 12 \, \phi^2  \, \Psi^{\prime2} + \Psi \, \Sigma_1 \right).
$$

Let us define the dimensionless fields $\tilde\phi=k^{\frac{2-d}{2}} \, \phi$,
$\tilde R=k^{-2} \, R$ and the dimensionless functions
$\tilde V(\tilde\phi^2)=k^{-d} \, V(\phi^2)$ and 
$\tilde F(\tilde\phi^2)=k^{2-d} \, F(\phi^2)$.
The beta functionals of the dimensionless and dimensionful functions are related as follows: 
\begin{eqnarray}
\label{eq:dlspotVFbeta1}
(\partial_t \tilde{V} )[ \tilde{\phi} ^{2}] &=& -d \, \tilde{V} (\tilde{\phi} ^2) + (d-2) \, \tilde{\phi} ^2 \, \tilde{V}'(\tilde{\phi} ^2) +  k^{-d} \, (\partial_t V)[\phi^2] \, \ ,\\
\label{eq:dlspotVFbeta2}
(\partial_t \tilde{F})[ \tilde{\phi} ^2] &=& -(d-2) \, \tilde{F} (\tilde{\phi} ^2) + (d-2) \, \tilde{\phi} ^2 \, \tilde{F}'(\tilde{\phi} ^2) +  k^{-(d-2)}\,  (\partial_t F)[\phi^2] \, \ .
\end{eqnarray}

Some comments are in order.
From the expressions of $\partial _t V$ and $\partial _tF$ given in eq. (\ref{eq:betaVF1}) and (\ref{eq:betaVF2})
respectively, we note that where ever there is occurence of $\phi^2$, it occurs in combinations
like $\phi^2 V' V'$, $\phi^2 V'F'$, $\phi^2 F'F'$, $\phi^2 V' \partial_t F'$, $\phi^2 F' \partial_t F'$, 
$\phi^2 V''$ and $\phi^2 F''$. Occurence of such combinations are crucial, as they help us 
(as is demostrated in \cite{gau_chris}) in proving that minimal coupling is self consitent.
Because of the occurrence of $\partial_t \tilde F'$ in the r.h.s. of \eqref{eq:betaVF1}, the system of equations
cannot be solved algebraically for $\partial_t\tilde F$.
It may be possible to solve it as a differential equation,
but here we shall not pursue this.
Rather, we observe that if $\tilde F$ and $\tilde V$ are assumed to be finite polynomials
in $\tilde\phi^2$ of the form
\begin{equation}
\label{eq:dimlesspotfinite}
\tilde{V} (\tilde{\phi} ^2) = 
\sum_{n=0}^{a} \tilde{\lambda}_{2n} \, \tilde{\phi}^{2n}\ ;\qquad
\tilde{F} (\tilde{\phi} ^2) = 
\sum_{n=0}^{b} \tilde{\xi}_{2n} \, \tilde{\phi}^{2n}\,.
\end{equation}
with finite $a$ and $b$, then  $\partial_t \tilde F'$ is also a finite polynomial in the beta functions
and it becomes possible to solve for the beta functions algebraically.
As an explicit example, in the appendix we give these equations in the de-Donder gauge
($\alpha=0$ and $\beta=1$) in $d=4$ with five couplings truncation ($a=2$, $b=1$).

\section{The Gaussian Matter Fixed Point}

\subsection{Minimal coupling is self consistent}
\label{sec:GMFP_mincoup}

We assume that $V$ and $F$ are real analytic so that they can be Taylor expanded
around $\phi^2=0$. 
A given $V$ and $F$ define a FP if the corresponding dimensionless potentials satisfy
$\partial_t\tilde V=0$ and $\partial_t\tilde F=0$.
Because of analyticity, this is equivalent to requiring that all the derivatives of 
$\partial_t\tilde V$ and $\partial_t\tilde F$ with respect to $\tilde\phi^2$, evaluated at
$\tilde\phi^2=0$ are zero.
Taking $n$ derivatives of eq. (\ref{eq:dlspotVFbeta1}) and eq. (\ref{eq:dlspotVFbeta2})
 with respect to $\tilde\phi^2$ we get
\begin{eqnarray}
\label{eq:important1}
0=(\partial_t\tilde{V})^{(n)}(0) &=&
((d-2)n-d) \, \tilde{V}^{(n)}(0)+(k^{-d} \, \partial_t V)^{(n)}(0) \, \ ;\\
\label{eq:important2}
0=(\partial_t\tilde{F})^{(n)}(0)&=&
(n-1)(d-2) \, \tilde{F}^{(n)}(0)+(k^{-(d-2)} \, \partial_t F)^{(n)}(0) \, \ .
\end{eqnarray}
where in the last two terms the expressions in brackets can be thought
of as functions of $\tilde\phi^2$.
We can rewrite them as
$$
\frac{\partial^n}{\partial(\tilde\phi^2)^n}(k^{-d}\partial_t V)
=k^{(d-2)n-d}\frac{\partial^n}{\partial(\phi^2)^n}(\partial_t V) \, \ ;\qquad
\frac{\partial^n}{\partial(\tilde\phi^2)^n}(k^{-(d-2)}\partial_t F)
=k^{(d-2)(n-1)}\frac{\partial^n}{\partial(\phi^2)^n}(\partial_t F) \, \ . 
$$

We now make the following Ansatz: 
\begin{gather}
V=k^d \, \tilde\lambda_0 \, \ , \qquad
F=k^{d-2} \, \tilde\xi_0 \, \ , 
\end{gather}
where $\tilde\lambda_0$ and $\tilde\xi_0$ are numbers to be determined.
This corresponds to putting $a=b=0$ in \eqref{eq:dimlesspotfinite},
or in other words to seeting to zero all scalar self couplings.
We are assuming here that all the derivatives of $V$ and $F$ at $\phi^2=0$ vanish,
so that $V$ and $F$ are just constants.
If a FP of this type exists, we call it a Gaussian Matter Fixed Point (GMFP).
In order to check that this ansatz defines a FP we need to show that eq. (\ref{eq:important1}) and 
(\ref{eq:important2}) 
are identically satisfied for all $n\geq1$, while for $n=0$ they
determine the numbers $\tilde\lambda_0$ and $\tilde\xi_0$.

For $n\geq1$ the first term on the r.h.s. of eq. (\ref{eq:important1}) and (\ref{eq:important2}) 
vanishes because of the ansatz. There remains to show that
$\frac{\partial^n}{\partial(\phi^2)^n}(\partial_t V)$
and $\frac{\partial^n}{\partial(\phi^2)^n}(\partial_t F)$
are zero at $\phi^2=0$.
In $d=4$ one can check this explicitly by inspecting eq. (\ref{eq:betaVF1}) and eq. (\ref{eq:betaVF2}).
The crucial point to observe is that in $\partial_t V$ and $\partial_t F$,
whenever $\phi^2$ appears explicitly, it is multiplied by some derivative of $V$ or $F$.
So when the derivative removes $\phi^2$, what remains is
zero because of the ansatz, and otherwise it is zero because there remains
some positive power of $\phi^2$.

In other dimensions this crucial property remains valid. 
In other dimensions this crucial property remains valid, because 
it is true either for the second variations (in the case of the transverse traceless tensor 
and transverse vector components) or for the matrix trace of the second variations,
in the case of the scalars.
Since the beta functionals are obtained by taking functional traces of these expressions,
this property will go through for them as well {\it i.e.} for
$n\geq1$ the eq. (\ref{eq:important1}) and (\ref{eq:important2})  are identically satisfied.
For a detailed proof see \cite{gau_chris}.  

Thus in any dimension the ansatz works for all $n\geq1$.
There remains to solve the equations for the constant
terms in $V$ and $F$, which are given by $\tilde\lambda_0$ and $\tilde\xi_0$.
We are going to do this numerically in section IV. In the meanwhile we assume that such a solution exists, 
and we study the properties of the linearized flow around it.

\subsection{Linearized Flow around GMFP}

To study the linearized flow around GMFP it will be convenient to Taylor expand $V$ and $F$ as follows:
\begin{equation}
\label{eq:potentals}
V(\phi^2)=\sum_{n=0}^{\infty}\lambda_{2n} \, \phi^{2n}\,\ ;\qquad
F(\phi^2)=\sum_{n=0}^{\infty}\xi_{2n} \, \phi^{2n}\,.
\end{equation}
We define dimensionless couplings $ \tilde{\lambda}_{2n} = k^{-d
+(d-2)n } \lambda_{2n}$ and $\tilde{\xi}_{2n}= k^{-(d-2)(1-n)} \xi_{2n} $, 
in such a way that the dimensionless potentials can be expanded as:
\begin{equation}
\label{eq:dimlesspot}
\tilde{V} (\tilde{\phi} ^2) = 
\sum_{n=0}^{\infty} \tilde{\lambda}_{2n} \, \tilde{\phi}^{2n}\ ;\qquad
\tilde{F} (\tilde{\phi} ^2) = 
\sum_{n=0}^{\infty} \tilde{\xi}_{2n} \, \tilde{\phi}^{2n}\, \ .
\end{equation}
To obtain the running of dimensionless couplings we take derivatives of eq. (\ref{eq:dlspotVFbeta1}) 
and eq. (\ref{eq:dlspotVFbeta2}) with respect to 
$\tilde\phi^2$ and use eq. (\ref{eq:betaVF_def}) 
\begin{equation}
\label{eq:dls_beta_cp}
\partial_t \tilde{\lambda}_{2n} = \left. \frac{1}{n!} \, \frac{ \delta ^n \partial_t \tilde{V}}{ \delta (\tilde{\phi}^2) ^n} \right |_{\tilde{\phi}^2=0}\, \ ; \qquad
\partial_t \tilde{\xi}_{2n} = \left. \frac{1}{n!} \, \frac{ \delta ^n \partial_t \tilde{F}}{ \delta (\tilde{\phi}^2) ^n} \right |_{\tilde{\phi}^2=0} \ .
\end{equation}
Because of the presence of $t$-derivative on the RHS of FRGE, we do not obtain the 
beta functions of dimensionless couplings directly,
rather we get algebraic equations for them, solving which one get the full beta functions. 

Having defined the dimensionless couplings, we now define the stability matrix 
to be the matrix of derivatives of the dimensionless beta functions with respect to the dimensionless 
couplings at the FP. By definition it is a tensor quantity in the theory space.
It will be convenient to write $V_0=V$ and $V_1 = -F$. One can then define the 
corresponding dimensionless potentials as $\tilde{V}_a = k^{d- 2 a} \, V_a$, where 
$a$ is either $0$ or $1$.
Then the stability matrix is given by,
\begin{equation}
\label{eq:stab_mat}
\left( M_{ij} \right) _{ab} = \left. \frac{ \delta \left ( \frac{1}{i!} \partial _t \tilde{V}_a^{(i)}(0) \right) }{ \delta  \left ( \frac{1}{j!} \tilde{V}_b^{(j)}(0)  \right)} \right |_{ FP} \, \ .
\end{equation}
From the above definition of the stability matrix we note that the couplings
get arranged in the following order:
$\lambda_0$, $\xi_0$, $\lambda_2$, $\xi_2$, $\lambda_4$, $\xi_4\ldots$.
Then the matrix $M$ at the GMFP has the following form:
\begin{equation}
\label{eq:stmatform}
\left(
\begin{array}{ccccc}
M_{00} & M_{01} & 0 & 0 & \cdots  \\
0 & M_{11} & M_{12} & 0 & \cdots \\
0 & 0 & M_{22} & M_{23} & \cdots \\
0 & 0 & 0 & M_{33} & \cdots \\
\cdots & \cdots &\cdots &\cdots & \cdots 
\end{array}
\right) \, \ ,
\end{equation}
where each entry is a $2 \times 2$ matrix of the form

\begin{equation}
\label{eq:Mij}
M_{ij}=
\left(
\begin{array}{c c}
\frac{\partial\beta^{\tilde{\lambda}}_{(2i)}}{\partial {\tilde{\lambda}}_{(2j)}}&
\frac{\partial\beta^ {\tilde{\lambda}}_{(2i)}}{\partial {\tilde{\xi}}_{(2j)}}\\
\frac{\partial\beta^{\tilde{\xi}}_{(2i)}}{\partial {\tilde{\lambda}}_{(2j)}}&
\frac{\partial\beta^{\tilde{\xi}}_{(2i)}}{\partial {\tilde{\xi}}_{(2j)}}
\end{array}
\right).
\end{equation}
Moreover the various non zero entries of $M$ are related to each other by the following 
recursion relations (in $d$-dimensions):
\begin{equation}
\label{eq:stab_mat_rel}
M_{ii} = (d-2) \, i + M_{00} \, \ ; \qquad
 M_{i,i+1} = (i+1) \, (2i + 1) \, M_{01} \, \ ,
\end{equation}
where
\begin{equation}
\label{eq:M00_M01}
M_{00}=
\left(
\begin{array}{c c}
-d & 0  \\
0 & -(d-2)  \\
\end{array}
\right)
+
\left(
\begin{array}{c c}
\delta M_{\tilde{\lambda}_0\tilde{\lambda}_0 }&
\delta M_{\tilde{\lambda}_0 \tilde{\xi}_0 }\\
\delta M_{\tilde{\xi}_0 \tilde{\lambda}_0 }&
\delta M_{\tilde{\xi}_0 \tilde{\xi}_0 }\\
\end{array}\right) \, \ ; \qquad
M_{01} = \left(
\begin{array}{c c}
\delta M_{\tilde{\lambda}_0\tilde{\lambda}_2 }&
\delta M_{\tilde{\lambda}_0 \tilde{\xi}_2 }\\
\delta M_{\tilde{\xi}_0 \tilde{\lambda}_2 }&
\delta M_{\tilde{\xi}_0 \tilde{\xi}_2 }\\
\end{array}\right) \, \ . 
\end{equation}

We can prove these facts for the one loop beta functions, {\it i.e.} neglecting the $t$-derivatives 
of the couplings on the r.h.s. of FRGE. 
Using this we note that the running of dimensionless potentials can be written as follows:
\begin{equation}
\label{eq:1loop_dls_run}
\partial _t \tilde{V}_a = - (d-2 a) \, \tilde{V}_a  + (d-2) \, \tilde{\phi}^2 \, \tilde{V}^{\prime}_a
+ \tilde H_a \left(\,\tilde{V}_b, \,\tilde{\phi}^2 \, \tilde V^\prime_b \tilde V^\prime_c, \, 
2  \tilde V^\prime_b + 4  \tilde\phi^2 \, \tilde V^{\prime\prime}_b \right) \, \ .
\end{equation}
We have indicated that the one loop beta functional depends on $\tilde\phi^2$ only through
the three types of combinations indicated as the arguments for $\tilde H_a$.
This can be verified in $d=4$ by inspection of eq. (\ref{eq:betaVF1}) and eq. (\ref{eq:betaVF2}),
 when one drops the terms proportional to $\partial_t F$ and $\partial_t F'$ in the r.h.s.
The properties of the stability matrix given above follow by taking successive derivatives of 
$\partial_t\tilde{V}_a $ with respect to $\tilde {\phi}^2$ at $\tilde {\phi}^2 =0$. 

The $i=0$ entries of eq. (\ref{eq:stab_mat}) can be calculated by setting $\tilde {\phi}^2 =0$ 
in eq. (\ref{eq:1loop_dls_run}):
\begin{equation}
\partial _t \tilde{V}_a(0)  = - (d-2 a) \, \tilde{V}_a(0)  
+\tilde{H}_a (\tilde{V}_a(0),\,2 \tilde{V}^{\prime}_a(0) ) \, \ .
\end{equation}
Since $\partial _t \tilde{V}_a(0)$ depends only on $\tilde{V}_a(0)$ 
and $\tilde{V}^{\prime}_a(0)$, in eq. (\ref{eq:stab_mat}) for $i=0$, only $j=0,1$ will be 
non zero. Thus $M_{00}$ and $M_{01}$ are given by,
\begin{equation}
\label{eq:M00_1loop}
\left( M_{00} \right)_{ab} = -(d-2 a) \, \delta_{ab} 
+ \left.\frac{\delta\tilde{H}_a(\tilde{V}_c(0),\,2 \tilde{V}^{\prime}_c(0))}{\delta \, \tilde{V}_b(0)} \right |_{GMFP} \, \ ; \qquad
\left( M_{01} \right)_{ab} =  \left. \frac{ \delta \tilde{H}_a (\tilde{V}_c(0), \,2  \tilde{V}^{\prime}_c(0)) }{ \delta \, \tilde{V}^{\prime}_b(0) } \right |_{GMFP} \, \ . 
\end{equation}

Now we take first derivative of $ \partial _t \tilde{V}_a $ with respect to $\tilde {\phi}^2$. This gives, 
\begin{eqnarray}
\label{eq:1_der_oneloop}
\partial _t \tilde V^{\prime}_a &=&  
-(d -2 a) \tilde V^{\prime}_a + (d-2)\tilde V^{\prime}_a  + (d-2) \tilde{\phi}^2  \tilde V^{\prime \prime}_a
+ \frac{\delta \tilde H_a}{\delta \tilde V_c} \tilde {V}^{\prime}_c 
+ \frac{\delta \tilde H_a}{\delta (\tilde{\phi}^2 \tilde V^{\prime}_c \tilde V^{\prime}_d )  } (  \tilde V^{\prime}_c \tilde V^{\prime}_d 
+ \tilde{\phi}^2  \tilde V^{\prime \prime}_c \tilde V^{\prime}_d +\tilde{\phi}^2  \tilde V^{\prime \prime}_d \tilde V^{\prime}_c )  \notag\\
&& 
+ \frac{\delta \tilde H_a}{\delta(2 \tilde V^{\prime}_c  + 4 \tilde{\phi}^2 \tilde V^{\prime \prime}_c)} 
(  2 \tilde V^{\prime\prime}_c  + 4 \tilde V^{\prime\prime}_c + 4 \tilde{\phi}^2 \tilde V^{\prime\prime\prime}_c ) 
\, \ .
\end{eqnarray}
When we set $\tilde{\phi}^2 =0 $, we note from the above equation 
that $\partial _t \tilde V^{\prime}_a(0) $ depends only on $\tilde V_a(0)$, $\tilde V^{\prime}_a(0 )$ 
and $\tilde V^{\prime\prime}_a (0)$. We use this in  eq. (\ref{eq:stab_mat}) to calculate the $i=1$ 
entries of the stability matrix.
We note that $ M_{1j} = 0$ for all $j \geq 3$. 
Now we find the remaining possible non zero entries. For $j=0$, we note that the dependence on $\tilde V_a(0)$ is present 
only in $\left. \frac{\delta \tilde H_a}{\delta \tilde V_c} \right |_{\tilde{\phi}^2 =0}$, 
$\left. \frac{\delta \tilde H_a}{\delta (\tilde{\phi}^2 \tilde V^{\prime}_c \tilde V^{\prime}_d )  } \right |_{\tilde{\phi}^2 =0}$ and 
$\left. \frac{\delta \tilde H_a}{\delta(2 \tilde V^{\prime}_c  + 4 \tilde{\phi}^2 \tilde V^{\prime \prime}_c)} \right |_{\tilde{\phi}^2 =0} $. 
But each of these terms are multiplied either with $\tilde V^{\prime}_a$ or $\tilde V^{\prime\prime}_a$, so when 
we calculate the stability matrix, 
these terms will not contribute due to GMFP conditions ($\tilde V^{(i)}_a = 0$ for all $i \geq 1$). 
Thus we conclude that $M_{10} = 0$.

For $j=1$, we take the derivative of $\partial _t \tilde V^{\prime}_a(0) $ with respect to $\tilde V^{\prime}_b$. 
Thus using the condition of GMFP and eq. (\ref{eq:M00_1loop}) we find,
\begin{equation}
\label{eq:M11_1loop}
\left( M_{11} \right)_{ab} = -(d-2 a )\delta_{ab} + (d-2) \delta_{ab} + \left. \frac{\delta \tilde H_a}{\delta \tilde V_b} \right |_{\tilde{\phi}^2 =0}
= (d-2) \delta_{ab} + \left( M_{00} \right)_{ab}  \, \ ,
\end{equation}
while for $j=2$ we take derivatve of $\partial _t \tilde V^{\prime}_a(0) $ with respect to 
$\tilde V^{\prime\prime}_b/2$ and use eq. (\ref{eq:M00_1loop}). Thus we get
\begin{equation}
\label{eq:M12_1loop}
\left( M_{12} \right)_{ab} = 2 \left. \frac{\delta \tilde H_a}{\delta (2 \tilde V^{\prime}_c ) } \right |_{\tilde{\phi}^2 =0} \cdot 6 \delta_{bc} = 6 \left( M_{01} \right)_{ab} \, \ .
\end{equation}
Thus we see that for $i=1$ we have,
\begin{equation}
M_{10}=0 \, \ ; \qquad M_{11} = (d-2) \cdot 1 + M_{00} \, \ ; \qquad
M_{12} = 2 \cdot 3 \, M_{01} \, \ ; \qquad M_{1j}=0 \, \ , \forall j \geq 3 \, \ .
\end{equation}

In order to understand the structure of the lines $i\geq2$ we will proceed by induction.
We assume that the $i$-th derivative has the following structure,
\begin{eqnarray}
\label{eq:assum}
(\partial _t \tilde V_a)^{(i)} &=& -(d- 2 a) \tilde V^{(i)}_a + (d-2) \left( \tilde{\phi}^2 \, \tilde V^{(i+1)}_a + i \, \tilde V^{(i)}_a \right) 
+ \left \{ \cdots +  \frac{\delta \tilde H_a}{\delta \tilde V_c} \tilde {V}^{(i)}_c 
+ \frac{\delta \tilde H_a}{\delta (\tilde{\phi}^2 \, \tilde V^{\prime}_c \, \tilde V^{\prime}_d )  } \left(\tilde{\phi}^2 \, \tilde V^{\prime}_c \, \tilde V^{\prime}_d \right)^{(i)} \right. \notag \\
&& + \left.  \frac{\delta \tilde H_a}{\delta(2 \, \tilde V^{\prime}_c  + 4 \tilde{\phi}^2 \, \tilde V^{\prime \prime}_c)} \left(2 \, \tilde V^{\prime}_c  + 4 \tilde{\phi}^2 \, \tilde V^{\prime \prime}_c \right)^{(i)} \right \} \, \ ,
\end{eqnarray}
where the ($\cdots$) denote expressions having at least two factors of derivatives of potentials,
which are irrelevant when calculating the entries of stability matrix.
Clearly this property is true for $i=1$.
We show that if it holds for a given value of $i$, then it also holds for $i+1$.
Thus we take one more derivative eq.(\ref{eq:assum}) and we find
\begin{eqnarray}
\label{eq:assum_der}
(\partial _t \tilde V_a)^{(i+1)} &=& -(d- 2 a) \tilde V^{(i+1)}_a + (d-2) \left( \tilde{\phi}^2 \, \tilde V^{(i+2)}_a + (i+1) \, \tilde V^{(i+1)}_a \right) 
+ \left \{ \cdots +  \frac{\delta \tilde H_a}{\delta \tilde V_c} \tilde {V}^{(i+1)}_c  + \frac{\delta^2 \tilde H_a}{\delta \tilde V_c \, \delta \tilde V_d}  \tilde {V}^{(i)}_c \tilde V^{\prime}_d \right. \notag \\
&& \left. 
+  \frac{\delta^2 \tilde H_a}{\delta \tilde V_e \, \delta (\tilde{\phi}^2 \, \tilde V^{\prime}_c \, \tilde V^{\prime}_d ) }  \tilde {V}^{(i)}_e \, \left(\tilde{\phi}^2 \, \tilde V^{\prime}_c \, \tilde V^{\prime}_d \right)^{\prime} 
+  \frac{\delta^2 \tilde H_a}{\delta \tilde V_e \, \delta (2 \, \tilde V^{\prime}_c  + 4 \tilde{\phi}^2 \, \tilde V^{\prime \prime}_c)}  \tilde {V}^{(i)}_e \,  \left(2 \, \tilde V^{\prime}_c  + 4 \tilde{\phi}^2 \, \tilde V^{\prime \prime}_c \right)^{\prime} \right. \notag \\
&& \left. 
+ \frac{\delta \tilde H_a}{\delta (\tilde{\phi}^2 \, \tilde V^{\prime}_c \, \tilde V^{\prime}_d )  } \left(\tilde{\phi}^2 \, \tilde V^{\prime}_c \, \tilde V^{\prime}_d \right)^{(i+1)}
+  \frac{\delta^2 \tilde H_a}{\delta \tilde V_e \, \delta (\tilde{\phi}^2 \, \tilde V^{\prime}_c \, \tilde V^{\prime}_d ) }  \tilde {V}^{\prime}_e \, \left(\tilde{\phi}^2 \, \tilde V^{\prime}_c \, \tilde V^{\prime}_d \right)^{(i)} \right. \notag \\
&& \left.
 +  \frac{\delta^2 \tilde H_a}{\delta (\tilde{\phi}^2 \, \tilde V^{\prime}_e \, \tilde V^{\prime}_f ) \, \delta (\tilde{\phi}^2 \, \tilde V^{\prime}_c \, \tilde V^{\prime}_d ) }  (\tilde{\phi}^2 \, \tilde V^{\prime}_e \, \tilde V^{\prime}_f )^{\prime} \, \left(\tilde{\phi}^2 \, \tilde V^{\prime}_c \, \tilde V^{\prime}_d \right)^{(i)} \right. \notag \\
&& \left. 
+  \frac{\delta^2 \tilde H_a}{\delta (2 \, \tilde V^{\prime}_e  + 4 \tilde{\phi}^2 \, \tilde V^{\prime \prime}_e ) \, \delta (\tilde{\phi}^2 \, \tilde V^{\prime}_c \, \tilde V^{\prime}_d ) } (2 \, \tilde V^{\prime}_e  + 4 \tilde{\phi}^2 \, \tilde V^{\prime \prime}_e)^{\prime} \, \left(\tilde{\phi}^2 \, \tilde V^{\prime}_c \, \tilde V^{\prime}_d \right)^{(i)}  \right. \notag \\
&& \left. 
+ \frac{\delta \tilde H_a}{\delta(2 \, \tilde V^{\prime}_c  + 4 \tilde{\phi}^2 \, \tilde V^{\prime \prime}_c)} \left(2 \, \tilde V^{\prime}_c  + 4 \tilde{\phi}^2 \, \tilde V^{\prime \prime}_c \right)^{(i+1)}
+  \frac{\delta^2 \tilde H_a}{\delta \tilde V_e \, \delta (2 \, \tilde V^{\prime}_c  + 4 \tilde{\phi}^2 \, \tilde V^{\prime \prime}_c)}  \tilde {V}^{\prime}_e \,  \left(2 \, \tilde V^{\prime}_c  + 4 \tilde{\phi}^2 \, \tilde V^{\prime \prime}_c \right)^{(i)} \right. \notag \\
&& \left. 
+  \frac{\delta^2 \tilde H_a}{\delta (2 \, \tilde V^{\prime}_e  + 4 \tilde{\phi}^2 \, \tilde V^{\prime \prime}_e ) \, \delta (\tilde{\phi}^2 \, \tilde V^{\prime}_c \, \tilde V^{\prime}_d ) } (2 \, \tilde V^{\prime}_e  + 4 \tilde{\phi}^2 \, \tilde V^{\prime \prime}_e)^{(i)} \, \left(\tilde{\phi}^2 \, \tilde V^{\prime}_c \, \tilde V^{\prime}_d \right)^{\prime} \right. \notag \\
&& \left.
+  \frac{\delta^2 \tilde H_a}{\delta (2 \, \tilde V^{\prime}_c  + 4 \tilde{\phi}^2 \, \tilde V^{\prime \prime}_c ) \, \delta (2 \, \tilde V^{\prime}_d  + 4 \tilde{\phi}^2 \, \tilde V^{\prime \prime}_d ) } (2 \, \tilde V^{\prime}_c  + 4 \tilde{\phi}^2 \, \tilde V^{\prime \prime}_c)^{(i)} \, (2 \, \tilde V^{\prime}_d  + 4 \tilde{\phi}^2 \, \tilde V^{\prime \prime}_d )^{\prime}
\right \} \, \ .
\end{eqnarray}
Aside from the new terms containing two factors of derivatives of the potentials,
which can be neglected for our purposes, the remaining terms have the same structure as 
eq. (\ref{eq:assum}). 
Thus by induction eq. (\ref{eq:assum}) holds for all $i$. 

We can now use this result to calculate the entries of the stability matrix in the $i$-th row. 
Using 
\begin{equation}
\label{eq:i_der}
\left(2 \, \tilde V^{\prime}_a  + 4 \tilde{\phi}^2 \, \tilde V^{\prime \prime}_a \right)^{(i)} = 2 \, (2 \, i +1) \, \tilde V^{(i+1)} _a 
+ 4 \tilde{\phi}^2 \, \tilde V^{(i+2)}_a \, \ ,
\end{equation}
we note that at $\tilde{\phi}^2 =0 $ and using the condition of GMFP for calculating the stability matrix we have,
\begin{equation}
M_{ij}=0 \, \ , \forall j \leq (i-1) \ ; \qquad M_{ii} = (d-2) \, i + M_{00} \, \ ; \qquad M_{i,i+1} = (i+1)\,(2 \, i +1) M_{01} \, \ ; \qquad M_{ij} = 0 \, \ , \forall j \geq (i+2) \, \ . 
\end{equation}
This completes the proof of our statements in the one loop approximation.
It is difficult to extend this proof to the exact equation, but we see in finite truncations
that the previous properties of the stability matrix remain true. 

Having established the properties of stability matrix we would like to compute its
eigenvalues. The good feature of the block structure of stability matrix indicated in eq. (\ref{eq:stmatform})
 is that the eigenvalues are given just by the diagonal blocks. Since the consecutive 
 diagonal blocks just differ by $d-2$, the eigenvalues of the consecutive diagonal blocks of
$M$ also differ by $d-2$. 
This is a very strong result, because it implies that, at a GMFP, the eigenvalues of $M$ are all
determined by the eigenvalues of $M_{00}$.
Furthermore, the off diagonal blocks of $M$ are all determined by $M_{01}$,
so knowing $M_{00}$ and $M_{01}$ one can also determine all the eigenvectors.
This is useful to understand the mixing among various operators at the FP.
The smallest truncation that is required to calculate both $M_{00}$ and $M_{01}$ is when we 
retain terms up to $\phi^2$ in each potential.

\section{Numerical Results}
\subsection{The GMFP in $d=4$.}
We now look for GMFP in various
dimensions and calculate the critical exponents of the system, which are defined 
to be the opposites of the eigenvalues of $M$, {\it i.e.}
$\theta_i = -\lambda_i $, where $\lambda_i$ is the eigenvalue. 
As explained in the previous section, it is enough to calculate the eigenvalues of $M_{00}$. 
We do this task first in $d=4$.

In $d=4$ for De-donder gauge we get the following FP equation,
\begin{eqnarray}
\label{eq:FPeq_d4}
24 \, \tilde\lambda_0 \left( 16 \pi^2 + \frac{1}{\tilde\lambda_0 - \tilde\xi_0} \right) &=& 19, \\
(265 - 2304 \, \pi^2 \, \tilde\xi_0) \, \tilde\xi_0 + ( 127 + 2304 \, \pi^2 \, \tilde\xi_0) \, \tilde\lambda_0 &=& \frac{160 \, \tilde\lambda_0 ^2}{\tilde\lambda_0-\tilde\xi_0} \, \ .
\end{eqnarray}
On solving these, the only real solution that we get is
\begin{gather}
\label{eq:FP}
\tilde{\lambda}^*_{0} = 0.00862 \,\ ; \qquad
\tilde{\xi}^*_{0} = 0.02375 \, \ .
\end{gather}
We now compute the critical exponents $\theta$ of the stability matrix in this gauge. 
The relations given in eq. (\ref{eq:stab_mat_rel}) 
between the various nonzero entries of the stability matrix are independent of the gauge. 
However the entires of $M_{00}$ and $M_{01}$ are gauge dependent.
For $d=4$ eq. (\ref{eq:stab_mat_rel}) reduces to,
\begin{gather}
\label{eq:stab_mat_rel_d4}
M_{ii} = 2 \, i + M_{00}\, \ ; \qquad
M_{i,i+1} = (i+1) \, (2 \, i + 1) \, M_{01}\, \ .
\end{gather}
In De-Donder gauge for $d=4$, the entries of $M_{00}$ are
\begin{align}
\label{eq:Mlam0lam0}
M_{\tilde{\lambda}_0 \tilde{\lambda}_0 } =& \frac{1}{32 \pi ^2}  \frac{1}{\Theta} \tilde{\xi}_0 
\Biggl[ 
\tilde{\xi}_0^2 \left\{ 9667 + 3456 \pi ^2 \tilde{\xi}_0 ( 169 + 2304 \pi ^2 \tilde{\xi}_0) \right\} + \tilde{\lambda}_0^2 \left\{ 3279 + 1152 \pi ^2 \tilde{\xi}_0 (275 + 6912 \pi ^2 \tilde{\xi}_0 ) \right\}  \notag \\
& - 18  \tilde{\lambda}_0 \tilde{\xi}_0 \left\{ 551 + 128 \pi ^2 \tilde{\xi}_0 ( 331 + 6912 \pi ^2 \tilde{\xi}_0 ) \right\} 
\Biggr] \, \ , \\[1mm] 
\label{eq:Mlam0xi0}
M_{\tilde{\lambda}_0 \tilde{\xi}_0 } =& \frac{1}{32 \pi ^2}  \frac{1}{\Theta} 
\Biggl[ 
-48384  \tilde{\lambda}_0 ^4 - 443520 \pi ^2 \tilde{\xi}_0^4 +\tilde{\lambda}_0  \tilde{\xi}_0^2 \left \{ -9667 + 2304 \pi ^2 \tilde{\xi}_0 ( 161 - 3456 \pi ^2 \tilde{\xi}_0 ) \right \}  \notag \\
&  - 3 \tilde{\lambda}_0 ^3 \left \{ 1093 + 4608 \pi ^2 \tilde{\xi}_0 ( 1 + 576 \pi ^2 \tilde{\xi}_0) \right \} + 18  \tilde{\lambda}_0 ^2 \tilde{\xi}_0 \left \{ 551 + 192 \pi ^2 \tilde{\xi}_0 ( -1 + 4608 \pi ^2 \tilde{\xi}_0 ) \right \} 
\Biggr] \, \ , \\[1mm]
\label{eq:Mxi0lam0}
M_{\tilde{\xi}_0 \tilde{\lambda}_0 } =& \frac{12}{\Theta} 
\Biggl[
  \tilde{\xi}_0^2 \left \{ -252  \tilde{\lambda}_0  \tilde{\xi}_0 ( 3 + 128 \pi ^2 \tilde{\xi}_0) + 3  \tilde{\lambda}_0 ^2 (41 + 1536 \pi ^2 \tilde{\xi}_0 ) +  \tilde{\xi}_0^2 ( 593 + 27648 \pi ^2 \tilde{\xi}_0 ) \right \} 
  \Biggr] \, \ , \\[1mm]
\label{eq:Mx0ixi0}
M_{\tilde{\xi}_0 \tilde{\xi}_0 } =& \frac{-3}{\Theta} 
\Biggl[
 -135 \tilde{\lambda}_0 ^4 + 1309 \tilde{\xi}_0^4 + 12  \tilde{\lambda}_0 ^3 \tilde{\xi}_0 ( 145 +1536 \pi ^2 \tilde{\xi}_0) + 36  \tilde{\lambda}_0 \tilde{\xi}_0 ^3 ( 77 + 3072\pi ^2 \tilde{\xi}_0 )  \notag \\ 
&  -6  \tilde{\lambda}_0 ^2 \tilde{\xi}_0^2 ( 811 + 23504 \pi ^2 \tilde{\xi}_0) \Biggr] \, \ .
\end{align}
While the entries of $M_{01}$ are
\begin{align}
\label{eq:Mlam0lam2}
M_{\tilde{\lambda}_0 \tilde{\lambda}_2 } =& \frac{1}{16 \pi ^2}  \frac{1}{\Theta} 
\Biggl[ \tilde{\xi}_0 ^2 ( 37 - 1152 \pi ^2 \tilde{\xi}_0) +  2 \tilde{\lambda}_0 \tilde{\xi}_0 ( -5 + 1152 \pi ^2 \tilde{\xi}_0 ) - \tilde{\lambda}_0^2  (7 + 1152 \pi ^2 \tilde{\xi}_0 )
 \Biggr] \, \ , \\[1mm]
\label{eq:Mlam0xi2}
M_{\tilde{\lambda}_0 \tilde{\xi}_2 } =& -\frac{3}{4 \pi ^2}  \frac{( 2  \tilde{\lambda}_0 - 5\tilde{\xi}_0  )(  \tilde{\lambda}_0 -\tilde{\xi}_0 )}{\Theta} \, \ , \\[1mm]
\label{eq:Mxi0lam2}
M_{\tilde{\xi}_0 \tilde{\lambda}_2 } =& \frac{24 \tilde{\xi}_0 (  \tilde{\lambda}_0 -\tilde{\xi}_0 )^2  }{\Theta} \, \ , \\[1mm]
\label{eq:Mxi0xi2}
M_{\tilde{\xi}_0 \tilde{\xi}_2 } =&  -\frac{72 \tilde{\xi}_0 (  \tilde{\lambda}_0 -\tilde{\xi}_0 )^2  }{\Theta} \, \ ,
\end{align}
where
\begin{equation}
\label{eq:del}
\Theta = \left[ -18 \tilde{\lambda}_0 \tilde{\xi}_0 ( 1+ 128 \pi ^2 \tilde{\xi}_0 ) + 3 \tilde{\lambda}_0 ^2 ( 5+ 384 \pi ^2 \tilde{\xi}_0 ) + \tilde{\xi}_0 ^2 ( -17 + 1152 \pi ^2 \tilde{\xi}_0) \right] ^2 \, \ .
\end{equation}

The relations eq. (\ref{eq:stab_mat_rel_d4}) tells that the critical exponents of consecutive diagonal blocks will differ by $2$. 
In the truncation where we
keep terms till $\phi^2$ in each potential, the critical exponents are,
\begin{gather}
\label{eq:eng_d4}
2.143 \pm 2.879 i \, \ , \qquad 0.143 \pm 2.879 i
\end{gather}
The critical exponents $ 2.143 \pm 2.879 i $ correspond to eigenvalues of $M_{00}$, while the
critical exponents  $ 0.143 \pm 2.879 i $ which are shifted by $2$ correspond to 
the eigenvalues of $M_{11}$. 
This justifies our claim. The eigenvectors in this truncation are
\begin{gather}
\label{eq:engvec_d4}
\left(
\begin{array}{c}
0.3557 \pm 0.3776 i \\
0.8549 \\
0 \\
0
\end{array}
\right) \, \ , \qquad
\left(
\begin{array}{c}
(-18.059 \pm 7.310 i) \times 10^{-4} \\
(-30.723 \pm 10.763 i)\times 10^{-4} \\
0.3557 \pm 0.3776 i \\
0.8549
\end{array}
\right) \, \ ,
\end{gather}
where the first complex conjugate pair of eigenvector correspond to critical exponents  $2.143 \pm 2.879 i $, 
while the second pair correspond to critical exponents $0.143 \pm 2.879 i $. 

\begin{figure}
[t!]\center
\resizebox{1\columnwidth}{!}
{\includegraphics{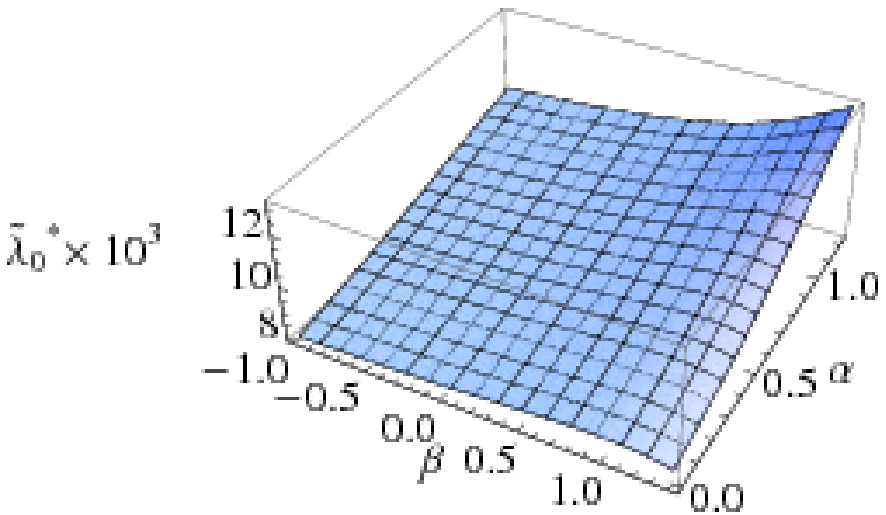} \includegraphics{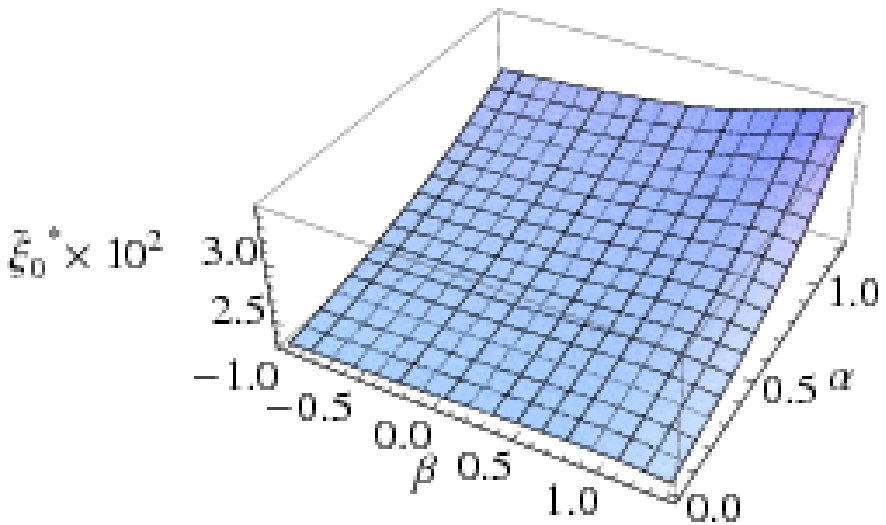} \includegraphics{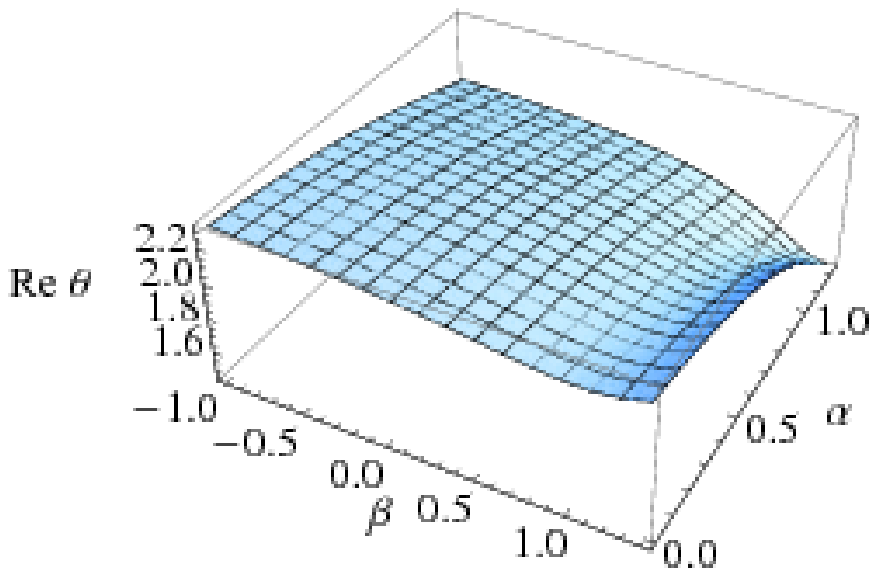}}
    \caption{\label{Fig:GMFP_th_gauge_d4}The gauge dependence of GMFP and critical exponents in $d=4$.
For convenience of depicting the values in plot, we have rescaled 
$\tilde\lambda^*_0$ and $\tilde\xi^*_0$ by $1000$ and $100$ respectively.}
\end{figure}

We then looked for GMFP in other gauges. In $d=4$ we consider various values of the gauge parameters 
$\alpha$ and $\beta$. 
To study the gauge dependence we considered $50$ different values of $\alpha$ 
in the range $0$ to $1.225$ at step of $0.025$,
and $25$ different values of $\beta$ in the range $-1$ to $1.4$ at interval of $0.1$. For each combination of 
$\alpha$ and $\beta$ we solved the FP equation obtained for $\tilde\lambda_0$ and $\tilde\xi_0$. 
In general, this produces a set of FPs. 
In order to choose the correct GMFP from that set, we plot all the real FPs to see which one is continously 
followed in other gauge values and which ones are spurious. For example one can take any 
value of $\beta$, and plot all the real FPs for various values of $\alpha$. 
Some FPs don't exits for all values of $\alpha$, and are assumed to be truncation artifacts.
Only one GMFP exists for all values, and is continuous. 
This observation of continuity in $\alpha$ and $\beta$ is useful to write 
a code for selecting the right GMFP for various gauges. 
After calculating the GMFP we calculate the critical exponents of $M_{00}$. 
We then plot the GMFP and critical exponents against the various gauge values and generate 3D graphs. 
In $d=4$ we obtain the graphs shown in Fig. (\ref{Fig:GMFP_th_gauge_d4}).
We note that the existence of the FP has been actually verified in a much larger range
of values of $\alpha$ and $\beta$.

\subsection{The GMFP in other Dimensions}
We now look for the GMFP in other dimensions. For any $d > 2$, in De-donder gauge, the FP equation for 
$\tilde\lambda_0$ and $\tilde\xi_0$ is given by,
\begin{eqnarray}
\label{eq:FPeq_d}
2 d  \tilde\lambda_0 + \frac{(3d -2)  \tilde\lambda_0 + (d-2)(d^2 + d-1 )  \tilde\xi_0}{(4 \pi)^{d/2} \Gamma \left(2 + \frac{d}{2}\right) ( \tilde\lambda_0 -  \tilde\xi_0) } &=& 0 \, \ , \\
\frac{A \, \tilde\lambda_0 ^2 + B \, \tilde\lambda_0 \tilde\xi_0 + C \, \tilde\xi_0^2}{( \tilde\lambda_0 -  \tilde\xi_0) }
+ D \, ( \tilde\lambda_0 -  \tilde\xi_0) &=& 0\, \ ,
\end{eqnarray}
where 
\begin{eqnarray}
A &=& -(d-1) (d^3 + 2 d^2 + 36 d +24) \, \ ,  \notag \\
B &=& -(d-1) (d^5 -17 d^3 -38 d^2 -96 d -48) \, \ , \notag \\
C &=& (d^6 -13 d^5 +32 d^4 -104 d^3 + 72 d^2 +36 d +24)\, \ , \notag \\
D &=& 24 d(d-1)(d-2) (4 \pi)^{d/2} \Gamma \left(2 + \frac{d}{2}\right) \, .
\end{eqnarray}
Solving these equations we find that in other dimensions, it is possible to have more than 
one real solution. 
But when we plot all the real the solutions against various $d$ in a graph, 
we notice that not all solutions exist in all dimensions. 
Only one solution exists in all dimensions, and is continuous in $d$. 
Besides, the ones which don't exist in all dimensions, have 
large critical exponents and are probably unphysical.
In Fig.2 we plot the position of the GMFP for $2<d\leq11$,
both in terms of $\tilde\lambda_0$ and $\tilde\xi_0$ and of the
more familiar dimensionless cosmological constant and Newton constant
\begin{equation}
\label{lambdag}
\tilde\lambda_0=\frac{2\tilde\Lambda}{16\pi\tilde G}\ ;\qquad
\tilde\xi_0=\frac{1}{16\pi\tilde G}\ .
\end{equation}

\begin{figure}
[t!]\center
\resizebox{1\columnwidth}{!}
{\includegraphics{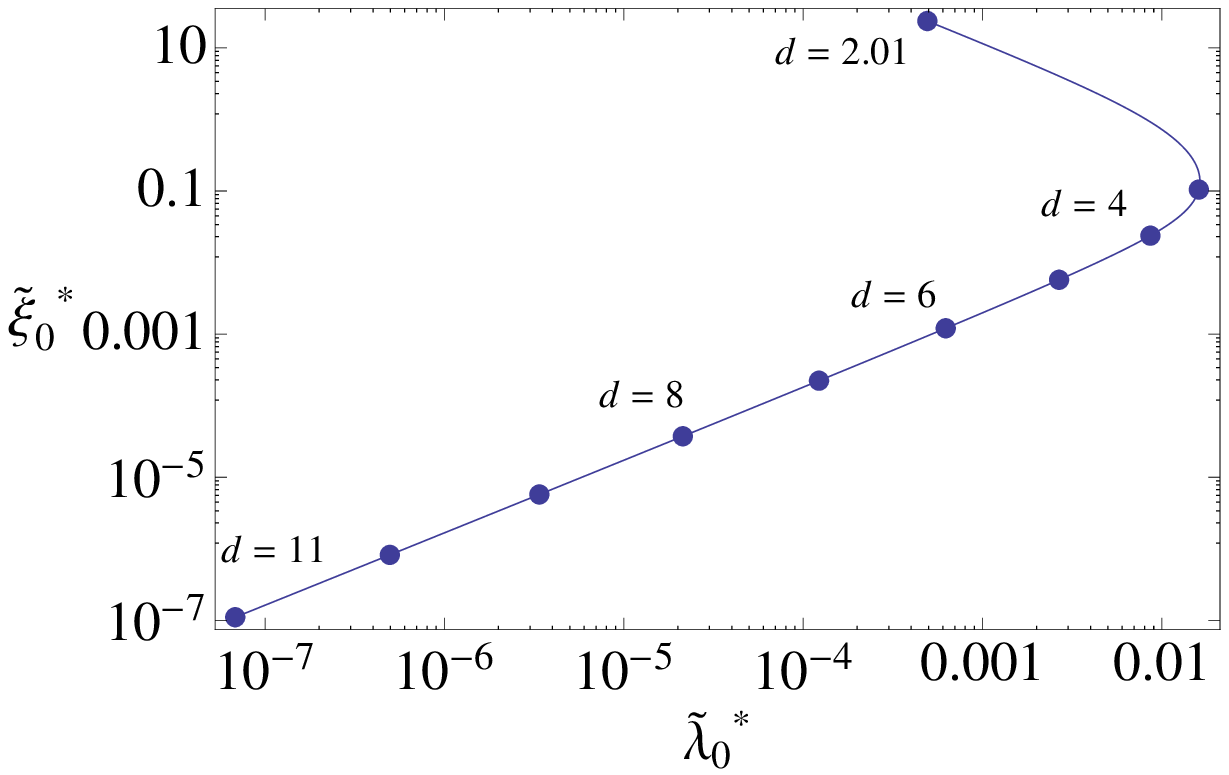} \includegraphics{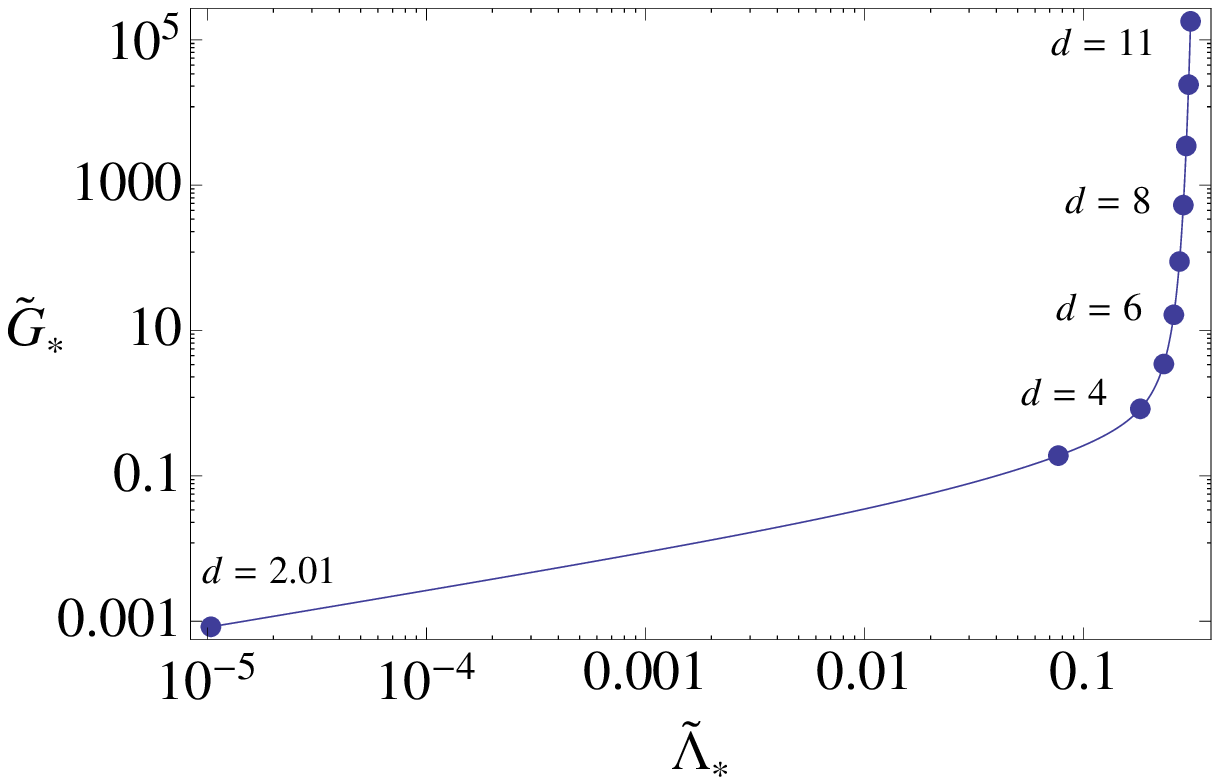}}
    \caption{\label{Fig:FPvsDim}
In the first graph we plot the GMFP $\tilde\lambda^*_0$ and $\tilde\xi^*_0$ in various dimensions. 
In the second plot we calculate the corresponding FP values of the 
cosmological constant $\tilde{\Lambda}^*$ and Newton's constant $\tilde{G}^*$ in various dimensions.}
\end{figure}
After having found the GMFP in various dimensions, we set to calculate their critical exponents.
In arbitrary dimensions, the various blocks of the stability matrix obey eq. (\ref{eq:stab_mat_rel}). 
We plot the critical exponents of $M_{00}$ for various dimensions.
\begin{figure}
[t!]\center
\resizebox{1\columnwidth}{!}
{\includegraphics{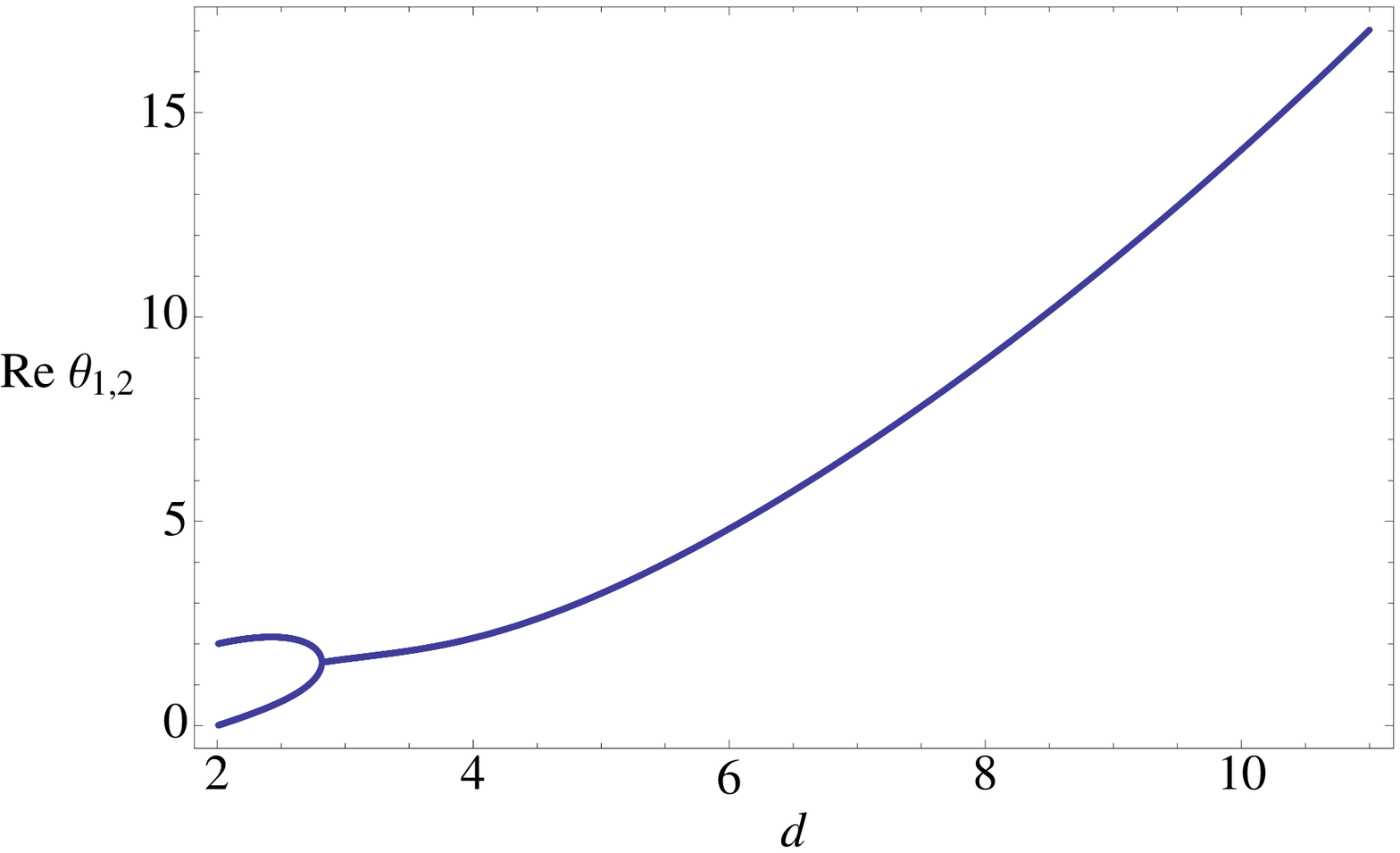} \includegraphics{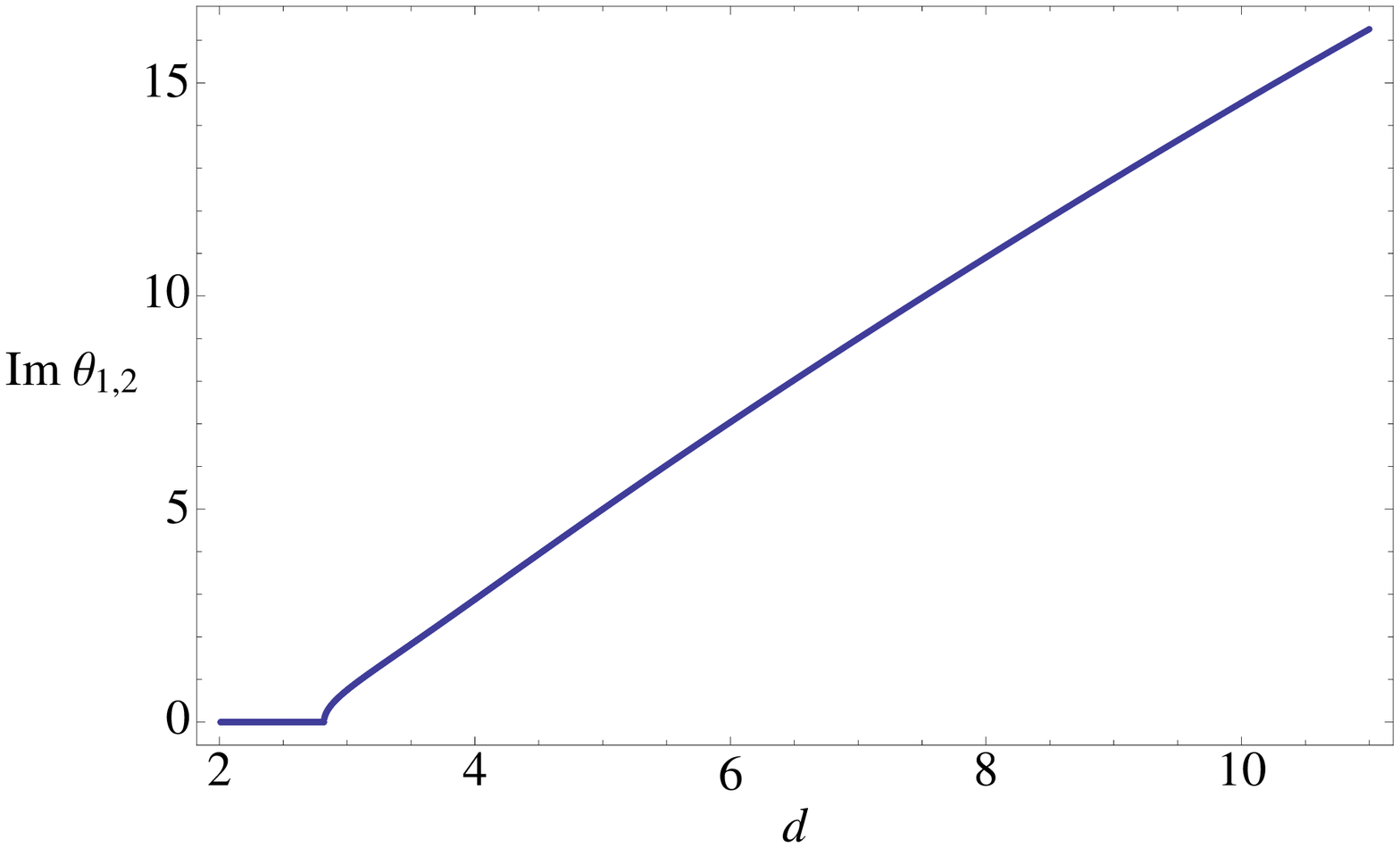}}
\caption{\label{Fig:EngvsDim}Critical exponents at the GMFP in various dimensions. 
The left panel shows the real part of the critical exponents,
the right panel shows the imaginary part of the critical exponents. 
We note that below $d=2.8$ the critical exponents becomes real.}
\end{figure}
From the graph Fig. (\ref{Fig:EngvsDim}) we note that around $d=2.8$ there is bifurcation.
 Below $d<2.8$ the critical exponents are no more complex.

A summary of the properties of the GMFP in various dimensions is given in table 
(\ref{tab:other_dim}). 
\begin{table}
\begin{center}
\begin{tabular}{|c|c|c|c|c|c|c|}
\hline
$d$ & $\tilde{\lambda}^*_0$ & $\tilde{\xi}^*_0$ & $\tilde{\Lambda}^*$
& $\tilde{G}^*$ & $\theta_1$ &  $\theta_2$ \\
\hline
2.001  &  4.968 $\times 10^{-5} $ & 2.386 $\times 10^2 $ & 1.041 $\times 10^{-7} $ & 8.339 $\times 10^{-5} $ & 2.001 & 0.001 \\
   3      &  1.605 $\times 10^{-2} $ & 1.047 $\times 10^{-1} $ & 7.666 $\times 10^{-2} $ & 1.900 $\times 10^{-1} $ & 1.627 + 0.754 $i$ &  1.627 - 0.754 $i$ \\
   4      & 8.620 $\times 10^{-3} $  & 2.375 $\times 10^{-2} $  & 1.814 $\times 10^{-1} $  & 8.375 $\times 10^{-1} $  & 2.143 + 2.879 $i$ & 2.143 - 2.879 $i$ \\
   5      & 2.669 $\times 10^{-3}$ & 5.744 $\times 10^{-3}$ & 2.323 $\times 10^{-1}$ & 3.463 & 3.236 + 4.996 $i$ & 3.236 - 4.996 $i$ \\
   6      &  6.230 $\times 10^{-4}$ & 1.207 $\times 10^{-3}$ & 2.581 $\times 10^{-1} $ & 1.648 $\times 10 $ & 4.818 + 7.039 $i$ &
                4.818 - 7.039 $i$ \\
   7      &  1.225 $\times 10^{-4}$ & 2.235 $\times 10^{-4}$ & 2.740 $\times 10^{-1}$ & 8.900 $\times 10$ & 6.744 + 9.004 $i$ &
               6.744 - 9.004 $i$ \\
   8      &  2.133 $\times 10^{-5}$ & 3.738 $\times 10^{-5}$ & 2.853 $\times 10^{-1}$ & 5.322 $\times 10^2$ & 8.945 + 10.904 $i$ 
           &  8.945 - 10.904 $i$ \\
   9      &  3.380 $\times 10^{-6}$ & 5.747 $\times 10^{-6}$ & 2.941 $\times 10^{-1}$ & 3.462 $\times 10^3$  & 11.396 + 12.748 $i$ 
           &  11.396 - 12.748 $i$ \\
  10     &  4.960 $\times 10^{-7}$ & 8.228 $\times 10^{-7}$ & 3.014 $\times 10^{-1}$ & 2.418 $\times 10^4$ &  14.089 + 14.537 $i$
           & 14.089 - 14.537 $i$ \\
  11     & 6.817 $\times 10^{-8}$ & 1.107 $\times 10^{-7}$ & 3.079 $\times 10^{-1}$ & 1.797 $\times 10^5$  & 17.025 + 16.261 $i$ 
           & 17.025 - 16.261 $i$ \\
\hline
\end{tabular}
\end{center}
\caption{Position of GMFP and critical exponents for various dimensions.}
\label{tab:other_dim}
\end{table}
Notice that for all the dimensions considered, the real part of the critical exponents
is greater than $d-2$ and less than $2(d-2)$. As a result, in all these cases there are exactly two pairs
of complex conjugate critical exponents with positive real part, {\it i.e.} four relevant directions.

\begin{figure}
[t!]\center
\resizebox{1\columnwidth}{!}
{\includegraphics{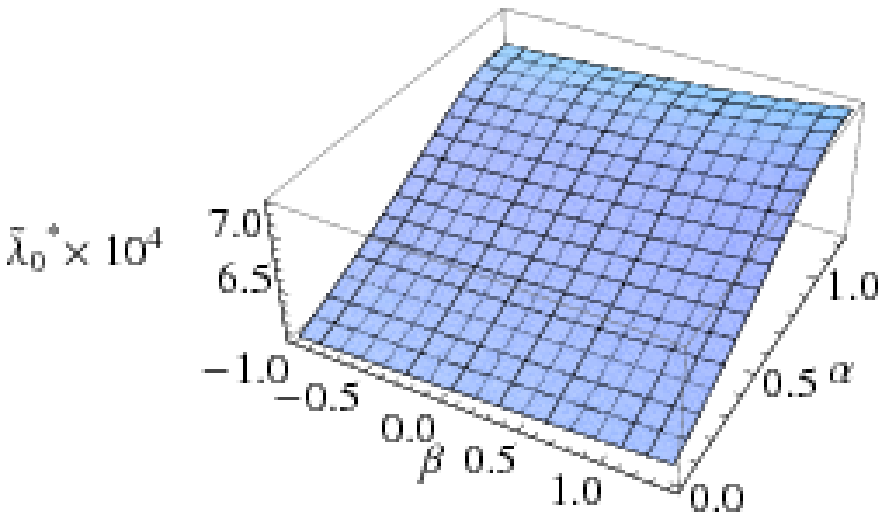} \includegraphics{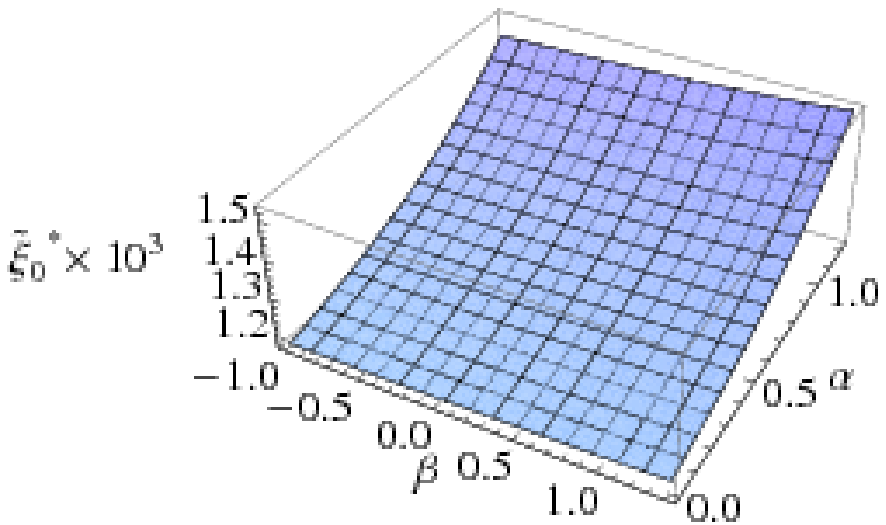} \includegraphics{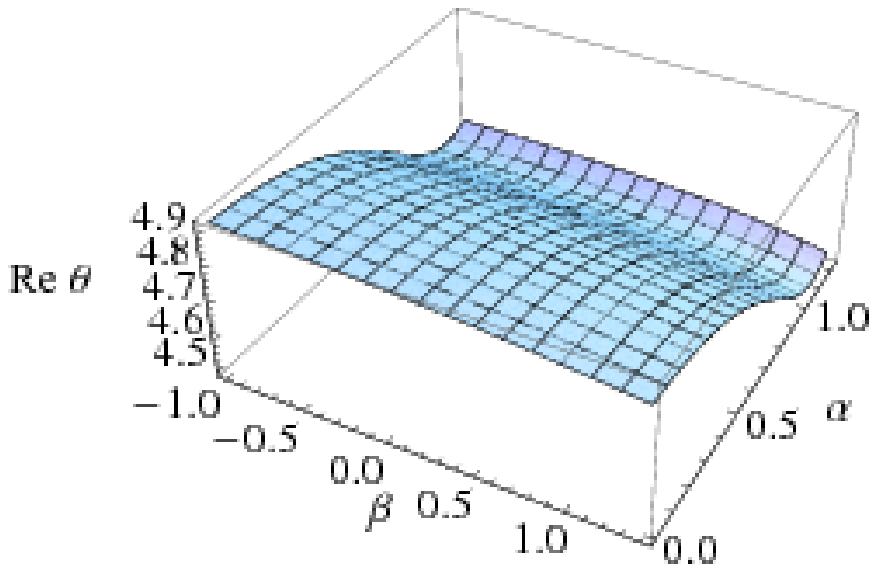}}
    \caption{\label{Fig:GMFP_th_gauge_d6}The gauge dependence of GMFP and critical exponents in $d=6$.
For convenience of depicting the values in plot, we have rescaled 
$\tilde\lambda^*_0$ and $\tilde\xi^*_0$ by $10^4$ and $10^3$ respectively.}
\end{figure}

Finally we studied the gauge dependence in different dimensions in the same wasy as we did in $d=4$, 
for example in $d=6$ we obtain the graphs shown in Fig. (\ref{Fig:GMFP_th_gauge_d6}).

\section{Other Non trivial Fixed Points}

Having discussed the existence and properties of the GMFP, we can ask ourselves
whether there exist other FP where the scalar field has nontrivial self-interactions.
We look for (truncated) polynomial FP potentials of the form
\begin{equation}
\tilde{V} (\tilde{\phi} ^2) = 
\sum_{n=0}^{a} \tilde{\lambda}_{2n} \, \tilde{\phi}^{2n}\ ;\qquad
\tilde{F} (\tilde{\phi} ^2) = 
\sum_{n=0}^{b} \tilde{\xi}_{2n} \, \tilde{\phi}^{2n}\,.
\end{equation}
with finite $a\geq 1$, $b\geq 0$.
Such potentials are known not to exist in a pure scalar theory in four dimensions \cite{hhm},
so we consider it unlikely that they exist in the presence of gravity,
In fact the outcome of our numerical searches is that no such FP's appear to exist in dimensions 4, 5 and 6.
(Some FP do appear in certain truncations but not in others, so they are likely to be just truncation artifacts.)

The situation is somewhat different in three dimensions.
We know that pure scalar theory in $d=3$ has the Wilson-Fisher FP \cite{wf}.
This FP can be seen in our calculations by taking the limit 
$\tilde G\to 0$ (where Newton's constant $G$ is related to $\xi_0=1/16\pi G$)
and $\tilde \lambda_0 \to 0$, in which case gravity decouples. 
Solving the FP equations of the scalar field in the 
LPA, truncated to order $\phi^4$, one gets $\tilde \lambda^*_2 = -0.0385$ and $\tilde \lambda^*_4 = 0.3234$, 
with critical exponents $\theta_1 = 1.843 $ and $\theta_2 = -1.176$.
(These are not very good values, but we quote them here for the sake of comparison
with what we find in the presence of gravity.)
The FP persists when one goes to higher truncations.

One wonders whether there exists a ``gravitationally dressed'' Wilson-Fisher FP,
with nonvanishing $\tilde G$, namely a FP where gravity and the scalar
simultaneously have nontrivial interactions.
Again in certain truncations one finds various FPs which turn out to be
truncation artifacts.
There seems however to exist one genuine FP: we find it in all truncations
where $a\geq b$, and it has very similar properties in all truncations.
To explore its properties we have looked in two directions:
increasing simultaneously $a$ and $b$, or keeping $b=0$ and increasing $a$.
\begin{table}
[h]
\begin{center}
\begin{tabular}{|c|c|c|c|c|c|c|c|c|c|c|}
\hline
$(a,b)$ & $\tilde{\lambda}^*_0$ & $\tilde{\lambda}^*_2$ & $\tilde{\lambda}^*_4$ & $\tilde{\lambda}^*_6$ & $\tilde{\lambda}^*_8$
       & $\tilde{\xi}^*_0$ & $\tilde{\xi}^*_2$ & $\tilde{\xi}^*_4$ & $\tilde{\xi}^*_6$ & $\tilde{\xi}^*_8$ \\
\hline
(2,1)   & 0.0196       &   -0.1646  &  -0.1595  &                    &                         
       & 0.1088       & -0.03108    &                   &                   &                         \\
(3,1)   & 0.01994     &  -0.1758    &  -0.1958 & -0.2796  &                         
       & 0.1096     &  -0.03810  &                   &                  &                           \\
(4,1)    & 0.02002     &  -0.1783   &  -0.2041  & -0.3466  & -0.5579       
        & 0.1098       & -0.03969    &                    &                   &                             \\
(2,2)    &  0.01894    &  -0.1408     & -0.1241    &                   &                     
        &  0.1071    & -0.01122    &  0.04297    &                   &                            \\
(3,2)    &   0.01971     &   -0.1680     &  -0.1848   & -0.2879  &                         
        &   0.1089   &    -0.03131    &  0.01731    &                   &                           \\
(4,2)    &  0.01988    &  -0.1735    &  -0.1975   &  -0.3544  &  -0.5687     
        &   0.1093   &  -0.03542    &  0.01121    &                    &                            \\
(3,3)    &    0.01911    &   -0.1469   &  -0.1469    &  -0.1935  &                        
        & 0.1074     &   -0.01420   &   0.05017     & 0.1617    &                         \\
(4,3)    &  0.01953    &  -0.1618    & -0.1768      & -0.3083    &   -0.6569    
       &  0.1084    & -0.02571        &  0.03197      & 0.1102     &                           \\
(4,4)     &   0.01923   &  -0.1512     &    -0.1572    &  -0.2496   &  -0.4911    
         &   0.1077    &  -0.01728    &    0.04732     & 0.1765     &  0.3868    \\
\hline
\end{tabular}
\end{center}
\caption{Position of Nontrivial FP in $d=3$ for various truncations.}
\label{tab:STFP_ot}
\end{table}
\begin{table}
[h]
\begin{center}
\begin{tabular}{|c|c|c|c|c|c|c|c|c|c|c|}
\hline
$(a,b)$ & $\theta_1^{\prime}$ & $\theta_1^{\prime \prime}$ & $\theta_3$ & $\theta_4$ & $\theta_5$
       & $\theta_6$ & $\theta_7$ &              &                 &                          \\
\hline
(2,1)    &   1.648        &     0.592       &    -0.956      &    -3.902        &     -13.46                    
        &                    &                     &                   &                     &                         \\
(3,1)   &    1.650         &    0.554         &  -1.079        &    -3.776        &    -11.20                     
       &    -29.397      &                      &                   &                     &                           \\
(4,1)   &    1.650          &    0.543         &    -1.105      &   -3.673         &     -10.02      
       &    -24.01       &    -49.31       &                     &                  &                             \\
\hline
$(a,b)$ & $\theta_1^{\prime}$ & $\theta_1^{\prime \prime}$ & $\theta_2^{\prime}$ & $\theta_2^{\prime \prime}$ 
       &        $\theta_5$          &                  $\theta_6$            &               $\theta_7$   &    $\theta_8$             &                 &                     \\
\hline
(2,2)    &    1.649     &      0.656      &     -7.979     &    1.261       &       -0.559              
        &     -3.192   &                     &                    &                    &                            \\
(3,2)    &   1.652      &     0.589       &     -7.933     &     3.909       &      -0.835                   
        &    -3.578    &  -27.67       &                    &                     &                           \\
(4,2)    &   1.652       &  0.570          &     -7.635      &  4.083         &      -0.898     
        &    -3.626   &   -22.64      & - 47.78      &                    &                            \\
\hline
$(a,b)$ & $\theta_1^{\prime}$ & $\theta_1^{\prime \prime}$ & $\theta_2^{\prime}$ & $\theta_2^{\prime \prime}$ 
       &  $\theta_3^{\prime}$ & $\theta_3^{\prime \prime}$ &               $\theta_7$   &    $\theta_8$             &   $\theta_9$          &                     \\
\hline
(3,3)    &   1.649     &    0.641       &     -6.703     &      2.097    &     -14.12                    
        &  8.990      &   -0.512       &   -2.991        &                   &                         \\
(4,3)    &   1.651      &  0.603         &     -6.448      &  3.343       &  -13.94     
        &   10.91   &    -0.657       &      -3.287     &  -42.28     &                           \\
\hline
$(a,b)$ & $\theta_1^{\prime}$ & $\theta_1^{\prime \prime}$ & $\theta_2^{\prime}$ & $\theta_2^{\prime \prime}$ 
       &  $\theta_3^{\prime}$ & $\theta_3^{\prime \prime}$ & $\theta_3^{\prime}$ & $\theta_3^{\prime \prime}$         &   $\theta_9$          &     $\theta_{10}$      \\
\hline
(4,4)    &    1.650     &      0.630    &      -5.958    &   2.008      &  -12.88    
        &     7.966    &     -20.07    &      19.03  &       -0.513  &  -2.977    \\
\hline
\end{tabular}
\end{center}
\caption{Critical exponents at Non trivial FP in $d=3$ for various truncations.
When the critical exponents are complex we write them in the form 
$\theta_\ell^\prime\pm i\theta_\ell^{\prime\prime}$}
\label{tab:STFP_eng_ot}
\end{table}

Tables (\ref{tab:STFP_ot}) and (\ref{tab:STFP_eng_ot}) give the position
and critical exponents of this FP for $a\geq b$ and $b\leq 4$.
One notices that $\tilde\lambda^*_{2n}<0$ for all $n>0$ in the table.
It is computationally demanding to continue in this direction,
so to have some indication on the sign of $\tilde\lambda^*_{2n}$
for higher $n$ we considered a simple truncation where $\tilde F$ is constant,
{\it i.e.} $\tilde\xi_n=0$ for $n>0$.
In this case we could push the truncation up to $a=8$.
The results are given in tables (\ref{tab:STFP_xi0}) and (\ref{tab:STFP_eng_xi0}).
One sees that the coefficients of the potential are indeed all negative.
%
%
\begin{table}
[h]
\begin{center}
\begin{tabular}{|c|c|c|c|c|c|c|c|c|c|c|}
\hline
$(a,b)$ & $\tilde{\lambda}^*_0$ & $\tilde{\lambda}^*_2$ & $\tilde{\lambda}^*_4$ & $\tilde{\lambda}^*_6$ & $\tilde{\lambda}^*_8$
& $\tilde{\lambda}^*_{10}$ & $\tilde{\lambda}^*_{12}$ & $\tilde{\lambda}^*_{14}$ & $\tilde{\lambda}^*_{16}$ & $\tilde{\xi}^*_0$ \\
\hline
(1,0)   &    0.01813     & -0.1088   &                   &                  &                 
       &                      &                  &                   &                  &    0.1060    \\
(2,0)   &     0.01880    & -0.1343  & -0.1561  &                  &                       
       &                        &                  &                   &                  &   0.1065     \\
(3,0)   &   0.01894     & -0.1395 & -0.1942  & -0.2633 &                        
       &                       &                  &                   &                   &    0.1066       \\
(4,0)   &   0.01898    &  -0.1407 & -0.2032  & -0.3284  &  -0.4998  
       &                      &                    &                   &                   &    0.1066       \\
(5,0)   & 0.01899       & -0.1410  &  -0.2053  &  -0.3437  &  -0.6182  
       & -0.9604      &                    &                    &                    &   0.1066        \\  
(6,0)   & 0.01899   &  -0.1411   &   -0.2058  &  -0.3472   & -0.6452    
       & -1.180   &  -1.826   &                    &                    &  0.1066         \\
(7,0)   & 0.01899   & -0.1411    &  -0.2059    &  -0.3479   &  -0.6511   
       & -1.228  &  -2.229   &  -3.380  &                    &  0.1066            \\
(8,0)   & 0.01899     &  -0.1411  &   -0.2059  &  -0.3481   & -0.6524    
       &    -1.238   &  -2.313  &  -4.091  & -5.977     &  0.1066        \\
\hline
\end{tabular}
\end{center}
\caption{Position of Nontrivial FP in $d=3$ for other truncations.}
\label{tab:STFP_xi0}
\end{table}
\begin{table}
[h]
\begin{center}
\begin{tabular}{|c|c|c|c|c|c|c|c|c|c|c|}
\hline
$(a,b)$ & $\theta_1^{\prime}$ & $\theta_1^{\prime \prime}$ & $\theta_3$ & $\theta_4$ & $\theta_5$
& $\theta_6$ & $\theta_7$ & $\theta_8$ & $\theta_9$ & $\theta_{10} $ \\
\hline
(1,0)   &    1.659      & 0.753        &   -1.699       &                  &                 
       &                   &                  &                   &                  &                    \\
(2,0)   &  1.675      &    0.745       &    -1.594      &   -10.59   &                       
       &                  &                    &                   &                 &                   \\
(3,0)   &  1.679      &   0.742        &   -1.485       &    -8.384      &    -22.41                    
       &                   &                   &                   &                   &                    \\
(4,0)   &     1.68      &  0.741        &    -1.434      &    -7.341      &     -18.19  
       &     -36.96   &                    &                   &                   &                     \\
(5,0)   &    1.68        &   0.741       &     -1.414     &    -6.84        &  -15.93 
       &    -31.12      &    -54.10    &                    &                  &                     \\  
(6,0)   &   1.68        &    0.741       &   -1.407       &  -6.609        &   -14.70   
       &  -27.47       &  -47.25       &    -73.60    &                    &                    \\
(7,0)   &    1.68      &    0.740       &    -1.405       &  -6.509        &  -14.02   
       & -25.287       &  -42.062       &  -66.663       &   -95.172     &                    \\
(8,0)   &   1.68      &    0.740       &    -1.405        &   -6.469       &   -13.67    
       & -23.94     & -38.77        &  -59.72        & -89.58       &  -118.4        \\
\hline
\end{tabular}
\end{center}
\caption{Critical Exponents at Non trivial FP in $d=3$ for other truncations. }
\label{tab:STFP_eng_xi0}
\end{table}
Furthermore, the coefficients grow in absolute value, so the series for $V$
has a very small radius of convergence.
This is similar to the situation discussed in \cite{hhm},
making the FP unphysical.
So we conclude that also in three dimensions there is probably
no physically viable FP besides the GMFP.

\goodbreak

\section{Conclusions}

The results of this paper confirm and extend the findings of \cite{Perini2}.
The GMFP is found to exist also in other dimensions and in other gauges,
and (with the possible exception of $d=3$) there does not seem to be other FP's 
with nontrivial scalar self-interactions.
In four dimensions this agrees with the findings of \cite{hhm}.

These results may be applied in various settings.
The beta functions given in Appendix A contain
the full dependence on the dimensionless parameters 
$\tilde\lambda_0$, $\tilde\xi_0$, $\tilde\lambda_2$, $\tilde\xi_2$,
$\tilde\lambda_4$, without making any assumption on the
value of these couplings (which in the case of the first three
means the ratio between the dimensionful couplings 
$\lambda_0$, $\xi_0$, $\lambda_2$ and the RG scale $k$).
In particular, threshold effects are taken into account
by the denominators $1+2\tilde\lambda_2$ and $\tilde\xi_0-\tilde\lambda_0$.
One can easily recognize among various terms the ones that are
obtained in perturbative approximations, but we emphasize that the 
derivation of these beta functions using the FRGE does not require
that the couplings be small.

The most natural application of these results seems to be in the context of early cosmology,
where a scalar field is used to drive inflation.
In an asymptotic safety context, it would be attractive to obtain inflation
as a result of FP behavior along the lines of \cite{Reutercosmology}.
In fact the energy scale involved is sufficiently high that one may expect
quantum gravity effects to play some role.
Alternatively, it would also be of interest to apply the flow equations
derived here to the scalar tensor theory, {\it e.g.} to improve the results
of \cite{higgsflaton}.

According to various speculations, quantum effects may play a role also
on very large scales, and then again the RG flow of scalar-tensor theory
could become relevant.
In this connection we recall that scalar-tensor theories of a different type
also arise in the conformal reduction of pure gravity,
and have been studied from the FRGE point of view in \cite{Weyer,crehroberto}.

We have mentioned in the Introduction that scalar-tensor theories can be
reformulated classically also as pure gravity theories with $f(R)$ type actions,
and one may wonder whether there is a relation also between their RG flows.
In particular one could ask whether the FP that was found in
\cite{CPR, MachSau} has a counterpart in the equivalent scalar-tensor theory.
At first sight one would think that this is not the case, because the choice 
of cutoff breaks the classical equivalence between these theories.
Still, this point deserves a more detailed investigation.

Another direction for research is the inclusion of other matter fields.
As discussed in the introduction, if asymptotic safety is indeed the answer 
to the UV issues of quantum field theory,
then it will not be enough to establish asymptotic safety of gravity:
one will have to establish asymptotic safety for a theory including
gravity as well as all the fields that occur in the standard model,
and perhaps even other ones that have not yet been discovered.
Ideally one would like to have a unified theory of all interactions
including gravity, perhaps a GraviGUT along the lines of \cite{nesti}.
More humbly one could start by studying the effect of gravity on
the interactions of the standard model or GUTs.
Fortunately, for some important parts of the standard model it is
already known that an UV Gaussian FP exists, so the question is whether
the coupling to gravity, or some other mechanism, can cure
the bad behavior of QED and of the Higgs sector.
That this might happen had been speculated long ago \cite{fradkintseytlin};
see also \cite{buchbinder} for some detailed calculations.
It seem that the existence of a GMFP for all matter interactions would be
the simplest solution to this issue.
In this picture of asymptotic safety, gravity would be the only effective interaction 
at sufficiently high scale.
The possibility of asymptotic safety in a nonlinearly realized 
scalar sector has been discussed in \cite{coza}.
Aside from scalar tensor theories, the effect of gravity has been studied
in \cite{Wilczek} for gauge couplings and \cite{Zanusso} for Yukawa couplings.

\appendix

\section{Explicit beta functions}
In section II.B we have presented equations which in principle determine the beta functionals 
for $V$ and $F$ in $d=4$ and in the gauge $\alpha=0$ and $\beta=1$.
At one loop the beta functionals can be immediately read off from there,
but if one wants to get the ``improved'' beta functionals (meaning that no approximation is made
beyond the truncation), then the system is too complicated to be solved.
It can be solved if we assume that $V$ and $F$ are finite polynomials.
We give here a set of linear equations that determine the beta functions in a 
five coupling truncation, including $\lambda_0$, $\xi_0$, $\lambda_2$, $\xi_2$, $\lambda_4$. 
These equations can be easily solved using algebraic manipulation software.
This exercise will also enable us to compare with familiar one loop results.
These beta functions had been written previously in \cite{Perini2}, but since there we had left the
cutoff generic, it was not possible to compute the integrals over momenta, 
which are contained in the expressions $Q_2$ and $Q_1$. 
\footnote{Note that the notation used in \cite{Perini2} is opposite to the one used here:
parameters with a tilde are dimensionful, those without tilde are dimensionless.
The beta functions written in \cite{Perini2} contain a number of transcription errors,
which however do not affect the subsequent results.}
Here the integrals have already been performed, using an optimized cutoff \cite{Litim:cf},
so the beta functions are in closed form and completely explicit. 
We use the notation 
$\eta=\frac{\partial_t\xi_0}{\xi_0}=\frac{\partial_t\tilde\xi_0}{\tilde\xi_0}+2$.

\begin{align}
\label{eq:beta.lam0}
\partial _t \tilde{\lambda}_0 = & -4 \tilde{\lambda}_0 
+ \frac{1}{32 \pi ^2} 
\left[2+\frac{1}{1+2\tilde{\lambda}_2}+ \frac{6\tilde{\lambda}_0}{\tilde{\xi}_0-\tilde{\lambda}_0} \right] 
+ \frac{\eta}{96 \pi ^2} \frac{5\tilde{\xi}_0-2\tilde\lambda_0}{\tilde{\xi}_0-\tilde{\lambda}_0}
, \\[1mm]
\label{eq:beta.xi0}
\partial _t \tilde{\xi}_0 = & -2 \tilde{\xi}_0 + \frac{1}{384 \pi ^2} 
\left[ 25 - \frac{4}{1 + 2  \tilde{\lambda}_2} - \frac{24  \tilde{\xi}_2}{(1+2\tilde{\lambda}_2)^2}
+ \frac{8 \tilde{\xi}_0 (7 \tilde{\xi}_0  - 2 \tilde{\lambda}_0 )}{(\tilde{\xi}_0-\tilde{\lambda}_0)^2} \right]
+ \frac{\eta}{1152 \pi ^2} 
\frac{17 \tilde{\xi}_0^2+18\tilde\xi_0\tilde\lambda_0-15\tilde\lambda_0^2}{(\tilde{\xi}_0-\tilde{\lambda}_0)^2}
, \\[1mm]
\label{eq:beta.lam2}
\partial _t \tilde{\lambda}_2 =& -2 \tilde{\lambda}_2 + \frac{1}{48 \pi ^2} 
\Biggl[ 
 \frac{9 \tilde{\lambda}_0 (1+ 2 \tilde{\xi}_2) }{2 (\tilde{\xi}_0-\tilde{\lambda}_0)^2} 
- \frac{9(2\tilde{\lambda}_0-\tilde{\xi}_0)  (1+ 2 \tilde{\xi}_2)^2 }{2 ( 1+ 2 \tilde{\lambda}_2 ) ( \tilde{\xi}_0-\tilde{\lambda}_0)^2}
- \frac{9 (1+ 2 \tilde{\xi}_2)^2 }{2 ( 1+ 2 \tilde{\lambda}_2 )^2  ( \tilde{\xi}_0-\tilde{\lambda}_0)} - \frac{18 \tilde{\lambda}_4 }
{( 1+ 2 \tilde{\lambda}_2 )^2}
\Biggr]
\notag\\
&+ \frac{\eta}{96 \pi ^2} 
\left[
- \frac{2 \tilde{\xi}_2}{\tilde{\xi}_0} + \frac{3 \tilde{\xi}_0 (1+ 2 \tilde{\xi}_2) }{2 ( \tilde{\xi}_0-\tilde{\lambda}_0)^2}
- \frac{3 \tilde{\xi}_0 (1+ 2 \tilde{\xi}_2)^2 }{2 ( 1+ 2 \tilde{\lambda}_2 ) ( \tilde{\xi}_0-\tilde{\lambda}_0)^2}
 \right] 
+  \frac{1}{96 \pi ^2} \frac{\partial_t \tilde{\xi}_2}{\tilde{\xi}_0} 
\left[ 
2 -\frac{3 \tilde{\xi}_0}{\tilde{\xi}_0-\tilde{\lambda}_0}
+\frac{6 \tilde{\xi}_0 (1+ 2 \tilde{\xi}_2) }{( 1+ 2 \tilde{\lambda}_2 ) ( \tilde{\xi}_0-\tilde{\lambda}_0)} 
\right], \\[1mm] 
\label{eq:beta.xi2}
\partial_t\tilde{\xi}_2 = & \frac{1}{576 \pi ^2}
\Biggl[
\frac{1+2\tilde{\lambda}_2}{\tilde{\xi}_0-\tilde{\lambda}_0} 
\left(9+\frac{39\tilde{\xi}_0}{\tilde{\xi}_0-\tilde{\lambda}_0}
+\frac{60 \tilde{\xi}_0^2}{(\tilde{\xi}_0-\tilde{\lambda}_0)^2} \right) 
 + \frac{3(3+32\tilde\xi_2)}{\tilde{\xi}_0-\tilde{\lambda}_0} 
- \frac{6\tilde{\xi}_0(11 + 2 \tilde{\xi}_2)}{(\tilde{\xi}_0-\tilde{\lambda}_0)^2} 
- \frac{60 \tilde{\xi}_0^2 (1 + 2  \tilde{\xi}_2)}{(\tilde{\xi}_0-\tilde{\lambda}_0)^3}
\notag\\
&
+\frac{216\tilde\xi_2(1+2\tilde{\xi}_2)^2}{(1+2\tilde{\lambda}_2)^3 (\tilde{\xi}_0-\tilde{\lambda}_0)}
+ \frac{9[\tilde{\lambda}_0(5-2\tilde{\xi}_2) - 2\tilde\xi_0(1 + 2 \tilde{\xi}_2)](1+2\tilde{\xi}_2)}{(1+2\tilde{\lambda}_2) (\tilde{\xi}_0-\tilde{\lambda}_0)^2} 
+ \frac{27 (1 + 2  \tilde{\xi}_2)(1-10\tilde{\xi}_2-16\tilde\xi_2^2)}{(1 + 2  \tilde{\lambda}_2)^2 (\tilde{\xi}_0-\tilde{\lambda}_0)}
\notag\\
&
+\frac{108\tilde\xi_0\tilde\xi_2(1+2\tilde{\xi}_2)^2}{(1+2\tilde{\lambda}_2)^2 (\tilde{\xi}_0-\tilde{\lambda}_0)^2}
+ \frac{72 \tilde{\lambda}_4 }{(1+2\tilde{\lambda}_2)^2}\frac{1+ 12\tilde{\xi}_2+2\tilde\lambda_2}{1+2\tilde{\lambda}_2}
\Biggr]
+ \frac{\eta}{1152 \pi ^2}  
\Biggl[ 
\frac{1 + 2  \tilde{\lambda}_2}{\tilde{\xi}_0-\tilde{\lambda}_0} 
\left(3 + \frac{18 \tilde{\xi}_0}{\tilde{\xi}_0-\tilde{\lambda}_0} + \frac{20\tilde{\xi}_0^2}{(\tilde{\xi}_0-\tilde{\lambda}_0)^2} \right) 
\notag \\
& 
+ \frac{15\tilde{\xi}_2}{\tilde{\xi}_0} 
- \frac{6(1+\tilde{\xi}_2)}{\tilde{\xi}_0-\tilde{\lambda}_0} 
-\frac{10 \tilde{\xi}_0(3+4\tilde{\xi}_2)}{(\tilde{\xi}_0-\tilde{\lambda}_0)^2} 
- \frac{20 \tilde{\xi}_0^2 (1 + 2  \tilde{\xi}_2)}{(\tilde{\xi}_0-\tilde{\lambda}_0)^3}
- \frac{3 [\tilde{\lambda}_0  -  \tilde{\xi}_0(5 - 4 \tilde{\xi}_2)] (1+2\tilde{\xi}_2)}{(1 + 2  \tilde{\lambda}_2) (\tilde{\xi}_0-\tilde{\lambda}_0)^2}
+ \frac{36  \tilde{\xi}_0\tilde\xi_2 (1 + 2  \tilde{\xi}_2)^2}{(1 + 2  \tilde{\lambda}_2)^2 (\tilde{\xi}_0-\tilde{\lambda}_0)^2}
\Biggr] 
\notag \\
&
+ \frac{1}{1152 \pi ^2} \frac{\partial_t\tilde{\xi}_2}{\tilde{\xi}_0} 
\left[
 -15+\frac{54 \tilde{\xi}_0 }{\tilde{\xi}_0-\tilde{\lambda}_0} + \frac{20 \tilde\xi_0 ^2}{(\tilde{\xi}_0-\tilde{\lambda}_0)^2}
- \frac{6 \tilde{\xi}_0(7+2 \tilde{\xi}_2)}{(1 + 2  \tilde{\lambda}_2) (\tilde{\xi}_0-\tilde{\lambda}_0)}
- \frac{144 \tilde{\xi}_0\tilde\xi_2 (1+2 \tilde{\xi}_2 )}{(1 + 2  \tilde{\lambda}_2)^2 (\tilde{\xi}_0-\tilde{\lambda}_0)}
 \right], \\[1mm]
\label{eq:beta.lam4}
\partial _t \tilde{\lambda}_4 =& \frac{1}{48 \pi ^2}
\Biggl[
\frac{9}{4 (\tilde{\xi}_0-\tilde{\lambda}_0)^2 }\Biggl(
5(1+2\tilde{\lambda}_2)(1+4\tilde{\xi}_2)
-(1+2\tilde{\xi}_2)(21+62\tilde{\xi}_2)  
+ \frac{33 (1 + 2  \tilde{\xi}_2)^3}{1 + 2  \tilde{\lambda}_2} 
\notag\\
&
- \frac{(1 + 2  \tilde{\xi}_2)^3 (23+24\tilde{\xi}_2)}{(1 + 2  \tilde{\lambda}_2)^2}
+ \frac{6 (1 + 2  \tilde{\xi}_2)^4}{(1 + 2  \tilde{\lambda}_2)^3}\Biggr)
+ \frac{9\tilde{\xi}_0(\tilde{\xi}_2-\tilde{\lambda}_2)^2}{(\tilde{\xi}_0-\tilde{\lambda}_0)^3} 
\left(6 \frac{(1+2\tilde{\xi}_2)^2}{(1 + 2  \tilde{\lambda}_2)^2} 
- 10\frac{1 + 2  \tilde{\xi}_2}{1 + 2  \tilde{\lambda}_2}+ 5 \right)
\notag\\
&-\frac{72\tilde{\lambda}_2\tilde\lambda_4(1+2\tilde\xi_2)(1 - 4\tilde{\lambda}_2 + 6\tilde\xi_2 )}
{(\tilde{\xi}_0-\tilde{\lambda}_0)(1+2\tilde\lambda_2)^3} 
+\frac{9\tilde{\xi}_0\tilde{\lambda}_4}{(\tilde{\xi}_0-\tilde{\lambda}_0)^2} 
\left(6 \frac{(1+2\tilde{\xi}_2)^2}{(1+2\tilde{\lambda}_2)^2} 
- 8 \frac{1 + 2  \tilde{\xi}_2}{1 + 2  \tilde{\lambda}_2}+ 3 \right)
+\frac{216 \tilde{\lambda}_4^2}{(1 + 2  \tilde{\lambda}_2)^3}  
\Biggr] 
\notag \\
& 
+ \frac{\eta}{96\pi^2}
\Biggl[
\frac{2\tilde{\xi}_2^2}{\tilde{\xi}_0^2} 
+\frac{3\tilde{\xi}_0(\tilde{\xi}_2-\tilde{\lambda}_2)^2}
{(\tilde{\xi}_0-\tilde{\lambda}_0)^3} 
\left(6\frac{(1 + 2  \tilde{\xi}_2)^2}{(1 + 2  \tilde{\lambda}_2)^2} 
- 10 \frac{1 + 2  \tilde{\xi}_2}{1 + 2  \tilde{\lambda}_2} + 5 \right) 
+ \frac{3 \tilde{\xi}_0 \tilde{\lambda}_4}{(\tilde\xi_0-\tilde\lambda_0)^2} 
\left(6\frac{(1+2\tilde{\xi}_2)^2}{(1+2\tilde{\lambda}_2)^2} 
- 8 \frac{1 + 2  \tilde{\xi}_2}{1 + 2  \tilde{\lambda}_2}+ 3 \right)
\Biggr]
 \notag\\
& 
+\frac{1}{96\pi^2}\frac{\partial_t\tilde{\xi}_2}{\tilde{\xi}_0} 
\Biggl[
-\frac{2\tilde{\xi}_2}{\tilde{\xi}_0} 
- \frac{3 \tilde{\xi}_0(\tilde{\xi}_2-\tilde{\lambda}_2)}
{(\tilde{\xi}_0-\tilde{\lambda}_0)^2} 
\left(12\frac{(1 + 2  \tilde{\xi}_2)^2}{(1+2\tilde{\lambda}_2)^2} 
- 21 \frac{1 + 2  \tilde{\xi}_2}{1 + 2  \tilde{\lambda}_2}+ 10 \right) 
- \frac{24\tilde{\xi}_0\tilde{\lambda}_4(1-4\tilde\lambda_2+6\tilde\xi_2)}{(1+2\tilde{\lambda}_2)^2(\tilde{\xi}_0-\tilde{\lambda}_0)}
\Biggr]\ .
\end{align}

If one neglects the terms involving $\eta$ and $\partial_t\tilde\xi_2$ in the r.h.s., 
then the remaining terms are the one loop beta functions for the couplings.
One can recognize among them some familiar terms.
The term containing $-18 \tilde\lambda_4$ in the first line of eq. (\ref{eq:beta.lam2}) and
the term containing $216 \tilde \lambda_4^2$ in the third line of eq. (\ref{eq:beta.lam4}) are the familiar
beta functions of the mass and of the coupling in $\phi^4$ theory in flat space.
The term containing $72\lambda_4(1+12\xi_2)$ in the third line of eq. (\ref{eq:beta.xi2})
is also known from earlier calculations \cite{buchbinder,higgsflaton}.

Notice the ubiquitous appearance of the factors $1/(1+2\tilde\lambda_2)$,
which represent threshold effects for the contributions of scalar loops:
for $k^2\gg \lambda_2$, $\tilde\lambda_2\ll 1$ and the denominator can
be approximated by $1$, whereas for 
$k^2\ll \lambda_2$, $\tilde\lambda_2\gg 1$ and the term is suppressed.
The denominators $\tilde\xi_0-\tilde\lambda_0$ have a somewhat similar effect.
When written in terms of the more familiar variables $\Lambda$ and $G$ 
defined in eq. (\ref{lambdag}), they give rise to denominators $(1-2\tilde\Lambda)$.
These can be approximated by $1$ when $\Lambda\ll k^2$
but they vanish when the dimensionless cosmological constant $\tilde\Lambda$
tends to $1/2$, corresponding to an infrared singularity in the RG trajectories.
This is well documented in the literature.

The term proportional to $9\tilde \xi_0 \tilde \lambda_4$ in the third line 
of eq. (\ref{eq:beta.lam4}) is the leading gravitational correction (of order $\tilde G$)
to the running of the scalar self coupling.
Note that for small $\tilde\lambda_0$, $\tilde\lambda_2$ and $\tilde\xi_2$,
the denominator and the bracket to its right can be expanded as
$1+O(\tilde\lambda_0)+O(\tilde\lambda_2)+O(\tilde\xi_2)$.
\footnote{on a related note, we also observe that the first term in $\partial_t\tilde\lambda_4$,
which is proportional to $\tilde G^2$, vanishes when we set $\tilde\lambda_2=0$
and $\tilde\xi_2=0$.}
The order of magnitude and sign of this term agree with the calculations done in \cite{GrigPerc},
in the gauge $\alpha=0$.
One should not expect the results to agree exactly, because this term is gauge dependent
and the calculation was done in a different gauge (namely $\beta=-1$).
Notice that this term is proportional to $\tilde \lambda_4$, 
thus when we set $\tilde \lambda_4$ to zero, the beta function 
for $\tilde \lambda_4$ does not get any contribution from gravity,
in agreement with the general statement that minimal coupling is self consistent. 
We observe that the same phenomenon happens in the case of the Yukawa coupling
\cite{Zanusso} and of the gauge coupling \cite{Wilczek}.



\end{document}